\documentclass{aa}  

\usepackage{graphicx}
\graphicspath{{figs/}}
\usepackage{txfonts}
\usepackage{ulem}
\usepackage[dvipsnames]{xcolor}
\usepackage{hyperref}
\hypersetup{
    colorlinks=true,       
    linkcolor=blue,          
    citecolor=blue,        
    filecolor=blue,      
    urlcolor=blue           
}
\usepackage[version=3]{mhchem}
\usepackage{lipsum} 
\usepackage{pifont}
\usepackage{fontawesome5}
\definecolor{orcidlogocol}{rgb}{0.65, 0.807, 0.223}
\newcommand{\orcid}[1]{$\,$\href{https://orcid.org/#1}{\textcolor{orcidlogocol}{\faOrcid}}}

\newcommand{\yes}{\ding{51}}
\newcommand{\Lsim}{{\Large $\sim$}}

\usepackage{placeins}

\begin{document} 

   \title{Anatomy of the Class I protostar L1489 IRS with NOEMA}
   \subtitle{I. Disk, streamers, outflow(s) and bubbles at 3mm\thanks{Datacubes are available at the CDS via anonymous ftp\linebreak to \url{cdsarc.u-strasbg.fr} (\url{130.79.128.5}) or via \url{https://cdsarc.cds.unistra.fr/viz-bin/qcat?J/A+A/}}} 

   \titlerunning{Anatomy of the Class I protostar L1489 IRS with NOEMA - I} 

   \author{M. Tanious\inst{1,2}\orcid{0009-0002-9761-8546}
          \and
          R. Le Gal\inst{1,2}\orcid{0000-0003-1837-3772}
          \and
          R. Neri\inst{2}\orcid{0000-0002-7176-4046}
          \and
          A. Faure\inst{1}\orcid{0000-0001-7199-2535}
          \and
          A. Gupta\inst{3}\orcid{0000-0002-9959-1933}
          \and
          C. J. Law\inst{4}\fnmsep\thanks{NASA Hubble Fellowship Program Sagan Fellow}\orcid{0000-0003-1413-1776}
          \and
          J. Huang \inst{5}\orcid{0000-0001-6947-6072}
          \and
          \\N. Cuello\inst{1}\orcid{0000-0003-3713-8073} 
          \and
          J. P. Williams\inst{6}\orcid{0000-0001-5058-695X}
          \and
          F. Ménard\inst{1}\orcid{0000-0002-1637-7393}
          }

    \institute{Université Grenoble Alpes, CNRS, IPAG, F-38000 Grenoble, France\\
            \email{
            maxime.tanious@univ-grenoble-alpes.fr, 
            romane.le-gal@univ-grenoble-alpes.fr
            }
        \and
            IRAM, 300 rue de la piscine, F-38406 Saint-Martin d’H\`{e}res, France
        \and
            European Southern Observatory, Karl-Schwarzschild-Str. 2, 85748 Garching bei München, Germany
        \and
            Department of Astronomy, University of Virginia, Charlottesville, VA 22904, USA
        \and
            Department of Astronomy, Columbia University, 538 W. 120th Street, Pupin Hall, New York, NY, United States of America
        \and 
            Institute for Astronomy, University of Hawaii, Honolulu, HI 96822, USA\\
             }

   \date{Received 29 November 2023 / Accepted 23 March 2024}

 
  \abstract
    {
    Over the past few years, chemical studies have revealed multiple structures in the vicinity of young stellar objects (YSOs). It has become evident that specific physical conditions are associated with the emission of particular molecular lines, allowing us to use molecular probes of the YSO physics. Consequently, chemical surveys are now necessary to fully constrain the origin of the observed structures. Several surveys have been conducted to explore the chemistry of YSOs, focusing on Class 0 and Class II objects. However, our knowledge of intermediate objects, that are Class I objects, remains limited.
    }
    {
    To bridge the gap and establish the relationship between observed structures and molecular line emission at the Class I evolutionary stage, we investigate the spatial distribution of key molecular gas species in the low-mass Class I protostar L1489 IRS (IRAS 04016+2610), a source part of the ChemYSO survey.
    }   
   {We performed a 3mm line survey at high spatial and high spectral resolution using the NOEMA interferometer and the IRAM-30m telescope. For the data analysis, we applied and compared two methods: a streamline model and the new python package \texttt{TIPSY}.}
   {We present here the ten brightest lines of our survey, in which we identified a new $\sim$\,3\,000~au long streamer in \ce{HC3N}, \ce{C2H}, and \ce{c-C3H2} emission, likely associated with more localized accretion shocks probed in SO. In addition, two $\sim$\,10\,000~au bubbles are seen with the dense molecular tracers \ce{HCO^+}, CS, and HCN around the YSO. We retrieve previously identified structures, like an outflow in \ce{HCO^+} and another streamer in \ce{C2H}.
   Additionally, potential indicators of a second outflow appear in CS and HCN emission, but its nature remains to be confirmed.}
   {
   The late infall identified at large scales may originate from the nearby prestellar core L1489 and is likely responsible for the formation of an external warped disk in this system. The detection of a potential second outflow could be the direct evidence of a binary system. Finally, we hypothesize that the bubbles may result from the magnetic pressure as observed in numerical simulations.
   }


   \maketitle
%

\section{Introduction}

Over the past decades, the latest generation of high-sensitivity telescopes has provided access to the cradles of star and planet formation at unprecedented spatial and spectral resolutions. They have uncovered countless structures at each stage of the planet formation, from the very early phases of young stars (Class 0) to planetary systems (Class III). Most of those structures can be probed using the gas seen with rotational transitions of molecules, which makes the study of chemistry of young stellar objects (YSOs) a powerful tool to investigate their physics. 

Through the past years, several programs investigated the chemical composition at the Class 0 stage -- for instance TIMASSS \citep{caux2011}, PILS \citep{jorgensen2016}, ASAI \citep{lefloch2018}, SOLIS \citep{ceccarelli2017}, GEMS \citep{fuente2019} -- or in the late protoplanetary disks or Class II stage -- such as DISCS \citep{oberg2010,oberg2011a}, CID \citep{guilloteau2016}, MAPS \citep{oberg2021}. 
However, much less is known about the chemistry of the intermediate evolutionary stage, that are Class I systems. The ChemYSO survey started to look at this chemistry in a more generic way, using IRAM-30m obervations of seven Class I YSOs in the Taurus star-forming region \citep{legal2020}. They appeared to be chemically rich with more than 20 molecules detected with less than three atoms in each source. However, interferometric observations are needed to access to their spatial distribution within the disks, to assess the impact of physical conditions on chemical evolution. 

The recent ALMA Large Program eDisk (PI: N. Ohashi) consisted in observations with a spatial resolution below 0\farcs1 of nineteen objects, including seven Class I YSOs \citep{ohashi2023}. Its main objective is determining whether substructures are present in Class 0/I disks \citep[e.g., in \object{Oph~IRS~63},][]{segura-cox2020}, but it offers in addition interesting results for linking chemistry to physics. For instance, it revealed the freeze-out of \ce{^12CO}, \ce{^13CO}, \ce{C^18O}, \ce{H2CO}, and SO near the midplane of the Class I object \object{IRAS~04302+2247}, whose CO snowline is located at 130~au \citep{lin2023}. Spiral infalling arms probed in SO and multiple streamers in \ce{C^18O} emission of \object{Oph~IRS~63} were also detected \citep{Flores2023}, as well as an expanding bubble in the CO~($2-1$) line emission in \object{Oph IRS 43} \citep{narayanan2023}.

Among the sources of the ChemYSO and eDisk surveys is \object{L1489 IRS} (also know as \object{IRAS 04016+2610}), an embedded Class I YSO located in the Taurus star-forming region \citep{gaia-collab2018}. It is a relatively isolated object still inside its parental molecular cloud Barnard 207 (B207) and at the edge of the prestellar core L1489. The position of the YSO suggests its potential migration from the core that would still feed material into the YSO’s disk \citep{brinch2007a}. 

\object{L1489 IRS} is known to have three nested disks: a 15~au radius inner disk, modeled by \citet{gramajo2010} from HST/NICMOS observations \citep{padgett1999}, and then observed in the continuum and SO emission with a position angle (P.A.) of 86\degr~\citep{yamato2023}. An intermediate disk, with a 200~au radius extent (P.A.~$=69\degr$), is traced in the continuum, \ce{^13CO} and \ce{C^18O} emission \citep{sai2020,yamato2023}. The latter is warped from a larger external Keplerian disk traced in \ce{C^18O} emission, and lying between 300~au and 600~au, with a P.A.~$=54\degr$~\citep{sai2020,yamato2023}. These three disks are embedded in an even larger molecular envelope \citep[$\sim$~5\,000~au,][]{sai2020} misaligned from the disk \citep{brinch2007b}. Some properties of this source from the literature are summarized in Table~\ref{table:KqL1489}. 

Previous observations at infrared and (sub)millimeters wavelengths suggested a potential unresolved binary system, which would explain the quadrupolar flow observed in \ce{H2}~$(1-0)$~S(1) emission, interpreted as two outflows \citep{lucas2000}, as well as the high luminosity of the object \citep{hogerheijde2000} despite its relative low mass \citep{brinch2007b}.
Additionally, infalling material within the envelope and a potential SO ring located at $\sim$ 300 au of the protostar(s) were reported by \citet{yen2014}. A dusty ring was also tentatively identified by \citet{ohashi2022} in 1mm continuum emission, subsequently confirmed by \citet{yamato2023}. 

As for the chemistry, several previous studies, which used single dishes to perform single pointing observations toward \object{L1489 IRS}, showed the molecular diversity of this object \citep{hirota2001,jorgensen2004,oberg2014,bergner2017,law2018,legal2020,mercimek2022}. Nevertheless, the low spatial resolution of these telescopes gives observed emission blending together all the structures of the object, resulting in an inability to infer where molecules are emitting from, and thus their associated physical conditions. 

Spatial distributions can only be accessed through interferometric observations for this source, according to its size. However, most of them focused on CO and its isotopologues so far \citep{hogerheijde1998,yen2013,yen2014,vanthoff2020,sai2020,sai2022,yamato2023}, resulting in a biased view of traced structures in this system. There is only few mappings of other lines but they are almost all 1mm observations \citep[with spatial resolutions from 8\arcsec\ to 0.1\arcsec, e.g.,][]{ohashi1996,hogerheijde2001,brinch2007b,yen2014,vanthoff2020,tychoniec2021,yamato2023}, limiting the accuracy of column density estimates through radiative transfer modeling, especially for sub-thermally excited species whose fundamental or lowest rotational transitions lie below 100~GHz. 

The chemical richness and the wide diversity of structures found in \object{L1489 IRS} make this source an ideal target to study the link between physical structures and molecular probes.
We present here the results of a 3mm mapping molecular survey conducted with the IRAM-30m and the NOrthern Extended Millimeter Array (NOEMA) at high spectral and high spatial resolution toward \object{L1489 IRS}. Section~\ref{sec:observations} describes the observations, the data reduction, and the imaging procedure. We report the observed structures in Sect.~\ref{sec:results}, especially infalling-like material for which we did an in-depth analysis in Sect~\ref{sec:analysis}. We discuss the inferred physical and chemical structure of \object{L1489 IRS} in Sect.~\ref{sec:discussion}, and summarize our results in Sect.~\ref{sec:conclusions}.

\begin{table}
\caption{Properties of \object{L1489 IRS} from the literature.}         
\label{table:KqL1489}      
\centering                          
\begin{tabular}{c c c }
\hline \hline
   Property & Value & Reference \\
\hline
   R.A. (J2000)  &  04:04:43.071  &  (1) \\[0.1cm]
   Dec. (J2000)  &  26:18:56.390  &  (1) \\[0.1cm]
   $T_\text{bol}$  &  213~K  &  (2) \\[0.1cm]
   $L_\star$  &  $3.4$~$L_\odot$  &  (2) \\[0.1cm]
   $M_\star$  &  $1.7\pm0.2$~$M_\odot$  &  (3) \\[0.1cm]
   $M_\text{Env.}$  &  $0.023_{-0.004}^{+0.010}\;M_\odot$  &  (4) \\[0.1cm]
   $M_\text{Disk}$  &  $0.009\pm0.001\;M_\odot$  &  (4) \\[0.1cm]
   $M_\text{Disk}/M_\text{Env.}$  &  0.39  & (4) \\[0.1cm]
   $R_\text{Env.}$  &  $\sim5\,000$~au  &  (5) \\[0.1cm]
   $R_\text{Disk}$  &  $\sim600$~au  & (5) \\[0.1cm]
   $v_\text{LSR}$  &  7.37~km~s$^{-1}$  &  (3) \\[0.1cm]
   $i$  &   $72$\degr     &  (3)  \\[0.05cm]
   Distance  &  $\sim$146~pc  &  (6) \\
\hline
\end{tabular}
\tablebib{(1) \citet{gaia-collab2018}, (2) \citet{ohashi2023}, \linebreak (3) \citet{yamato2023}, (4) \citet{sheehan2017}, (5) \citet{sai2020}, (6) \citet{roccatagliata2020}.}
\end{table}

\section{Observations}
\label{sec:observations}

\subsection{NOEMA observations}

\begin{table*}[!ht]
\caption{Primary targeted lines.}             
\label{table:lines+intensities}      
\centering                          
\begin{tabular}{l c c c c c c}        
\hline\hline                 
    Transition & $E_\text{up}$ & Rest Frequency & $\theta_\text{maj} \times \theta_\text{min}$ (PA) & Per-channel rms\tablefootmark{\dag} & Velocity Range \tablefootmark{*} & Integrated intensity \\
    & (K) & (GHz) & (\arcsec $\times$ \arcsec, \degr) & (mJy~beam$^{-1}$) & (km~s$^{-1}$) & (mJy~km~s$^{-1}$) \\
\hline    
    \ce{HCO+}~($1-0$)                    & 4.3  & 89.188525 & $2.05\times1.43$ (23.1) & 1.48 & [$-7.89$, 20.75] & $\geq112\,785$ \tablefootmark{a} \\
    \ce{H^13CO+}~($1-0$)                 & 4.2  & 86.754288 & $2.13\times1.48$ (22.5) & 1.47 & [4.01, 10.79]    & $\geq22\,315$ \tablefootmark{a} \\
\hline 
    \ce{HCN}~($1-0$)                     & 4.3  & 88.631602 & $2.10\times1.46$ (22.9) & 1.60 & [$-5.60$, 18.54] & $\geq86\,295$ \tablefootmark{a,b} \\
    \ce{H^13CN}~($1-0$)                  & 4.1  & 86.339921 & $2.16\times1.49$ (21.6) & 1.59 & [$-3.24$, 17.76] & $230\pm23$\tablefootmark{b} \\
    \ce{HC3N}~($11-10$)                  & 28.8 & 100.07639 & $1.82\times1.30$ (24.4) & 1.45 & [4.75, 9.49]     & $\geq7\,210$ \tablefootmark{a} \\
\hline 
    \ce{C2H}~($1_{1.5,2}-0_{0.5,1}$)         & 4.2  & 87.316898 & $2.12\times1.47$ (22.4) & 1.39 & [5.01, 9.72]     & $\geq21\,135$ \tablefootmark{a,c}\\
    \ce{c}-\ce{C3H2}~($2_{0,2}-1_{1,1}$) & 6.4  & 82.093544 & $2.31\times1.58$ (22.1) & 1.68 & [5.91, 8.10]     & $\geq9\,590$ \tablefootmark{a} \\
\hline
    \ce{CS}~($2-1$)                      & 7.1  & 97.980953 & $1.85\times1.32$ (24.6) & 1.44 & [4.44, 10.65]    & $\geq49470$ \tablefootmark{a} \\
    \ce{SO}~($2_3-1_2$)                  & 9.2  & 99.299870 & $1.83\times1.31$ (24.5) & 1.42 & [3.45, 10.77]    & $\geq20\,770$ \tablefootmark{a} \\
    \ce{SO}~($2_2-1_1$)                  & 19.3 & 86.093950 & $2.17\times1.49$ (21.6) & 1.45 & [4.25, 10.87]    & $156\pm16$ \\
\hline                                   
\end{tabular}
\tablefoot{$E_\text{up}$ and rest frequencies are from the CDMS \citep{muller2001,muller2005}.\\ 
\tablefoottext{\dag}{Estimated from line-free channels of cubes before primary beam correction.} \\
\tablefoottext{*}{LSR velocity range over which moment maps are produced and the integrated intensity derived.} \\
\tablefoottext{a}{As the emission is more extended than the primary beam, a lower limit is given.} \\
\tablefoottext{b}{The line properties are those of the major hyperfine component. All hyperfine components are integrated together for the flux.} \\
\tablefoottext{c}{While six hyperfine components are detected, we only present the brightest one here.}}
\end{table*}

\subsubsection{Description of the observations}
Interferometric observations of
\object{L1489 IRS} were carried out with NOEMA in Band 1 (Project IDs: S20AH and W20AJ, PI: R. Le Gal). The phase tracking center of this single pointing observations was $\alpha$(J2000) = 04\textsuperscript{h}04\textsuperscript{m}43\fs071, $\delta$(J2000) = 26\degr18\arcmin56\farcs390. The observations combine C- and A-array configurations from November 10, 2020 to March 25, 2021, with projected baselines ranging from 17.4~m (5.2~k$\lambda$) to 759.8~m (226~k$\lambda$). Observations in the C configuration were conducted, first, with 10 antennas on November 10, 15, and 16, 2020, and then with 11 antennas on March 20 and 25, 2021, resulting in an on-source time of 9.3~h. Observations in A configuration utilizing 11 antennas  were performed on February 24 and 26, 2021, accumulating an on-source time of 6.7~h. These observations, combining A- and C-array configuration, yield a synthesized beam $\sim$\,1\farcs4~at 90~GHz using natural weighting. The primary beam of the NOEMA antennas is $\sim$\,55\arcsec~at $\sim$\,90~GHz. 
For both projects (S20AH and W20AJ), \object{3C~84} served as the bandpass calibrator, and \object{MWC~349} and \object{LKH$\alpha$~101} served as flux calibrators. The uncertainty for the derived intensities are below 10\%. \object{QSO~B0400+258} and \object{4C~32.14} served as phase and amplitude calibrators for S20AH, \object{0354+231} and \object{4C~32.14} served as phase and amplitude calibrators for W20AJ. 

NOEMA's correlator covered a total instantaneous nominal bandwidth of $\sim$15.5~GHz per polarization, from 81.9 to 89.6~GHz and 97.4 to 105.1~GHz, with a resolution of 2~MHz.
Within these frequency ranges, the correlator was set up to provide in parallel 128 high-resolution windows, each 64~MHz wide and with an effective spectral resolution of 62.5~kHz, corresponding to velocity resolutions of $\sim$\,0.2 km~s$^{-1}$.
Table~\ref{table:lines+intensities} shows the brightest lines identified in our survey and used for the present study. Other molecular lines were also covered in this dataset but their emission is significantly weaker. Consequently, the analysis and presentation of these lines will be the focus of a forthcoming study.

\subsubsection{Data calibration}
The data calibration was performed using the NOEMA pipeline in the Continuum and Line Interferometer Calibration (\texttt{CLIC}) software, which is part of the Grenoble Image and Line Data Analysis Software (\texttt{GILDAS}\footnote{\url{http://www.iram.fr/IRAMFR/GILDAS}}) distribution.
Visibilities exceeding 1\farcs5 or 1\farcs6 seeing threshold in the C or A configuration, respectively, or surpassing an atmopsheric phase rms of 80\degr were flagged and excluded from the analysis due to a significant loss of data quality.
Calibrated visibilities were then stored in spectral window specific \textit{uv} tables for both high and low resolutions. 
Subsequent data processing were then realized with the \texttt{GILDAS MAPPING} program. 

\subsubsection{Data reduction}
For the continuum data, \textit{uv} continuum tables were created for both the lower sideband (LSB) and the upper sideband (USB) of the correlator, by applying a 3$\sigma$ threshold to filter out line emission in the 2~MHz tables using the \texttt{uv\_filter} and \texttt{uv\_average} commands. Three iterations of phase self-calibration were performed on these continuum tables using the \texttt{selfcal} task with masks defined around the $3\sigma$ emission. We applied 45s solution intervals on 100, 400 and 800 iterations cycle, using visibilities having a signal to noise ratio (S/N) above 3. Visibilities with an $\text{S/N}<3$ were retained after the process to enhance the S/N of the continuum image and preserve the synthesized beam size. The achieved noises are 5.70\,$\mu$Jy~beam$^{-1}$ for the LSB (i.e., 1.03 times the expected thermal noise), and 7.32\,$\mu$Jy~beam$^{-1}$ for the USB (i.e., 1.20 times the expected thermal noise). The gains estimated for each sideband from the selfcalibration of its continuum table were then applied to its lines-filtered table using the \texttt{uv\_cal} task. The LSB and the USB selfcalibrated lines-filtered tables were then merged using the \texttt{uv\_merge} task with a spectral index $\alpha=2.295\pm0.055$ (estimated from those tables). Finally, a continuum table of this merging was created with \texttt{uv\_average}.

Line-specific \textit{uv} tables were produced from the 62.5~kHz spectral resolution windows,
by extracting a 20~MHz spectral window centered on the line's rest frequency in the \textit{uv} plane (using command \texttt{uv\_extract}), followed by a subtraction of a 1st order baseline obtained from line-free emission (with the \texttt{uv\_baseline} command).
Notably, gain solutions from the self-calibrated continuum (LSB or USB, depending on the line's frequency) were not applied to the line \textit{uv} tables, as they did not significantly improve the S/N.

\subsubsection{Data imaging}
The \textit{uv} tables were imaged with natural weighting and then deconvolved using the Högbom CLEAN algorithm \citep{hogbom1974} by specifying a stopping criterion on the maximal intensity in the residual image. For the continuum, we first performed a shallow clean using three times the expected thermal noise to identify the emission. Subsequently, we defined a mask and cleaned down to 0.5 times the expected thermal noise to reach convergence. After using the \texttt{uv\_restore} command, we obtained the 3.2~mm dust continuum image shown in Fig.~\ref{fig:continuum}. As for the lines, we performed a cleaning down to three times the expected thermal noise without mask. The continuum peak was found to be offset from the phase tracking center. The tables have thus been reprojected on the continuum peak localized in $\alpha$(J2000)~=~04\textsuperscript{h}04\textsuperscript{m}43\fs085, $\delta$(J2000)~=~26\degr18\arcmin56\farcs206.

\subsection{IRAM-30m observations}

Single dish observations of \object{L1489 IRS} were carried out with the IRAM-30m between July 21 and 26, 2021 for a total on-source time of 29.4h (Project ID: 184-20, PI : R. Le Gal).
The Eight MIxer Receiver (EMIR) E090 was used and connected to the narrow fast Fourier Transform Spectrometers (FTS~50) backends, offering $4 \times 1.82$~GHz bandwidth at 50~kHz channel resolution ($\sim$\,0.2~km~s$^{-1}$) per spectral setup. Four spectral setups were required to encompass the entire bandwidth covered by the NOEMA observations. 
A 3\arcmin $\times$\,3\arcmin\, region centered on \object{L1489 IRS} was mapped using the On-The-Fly position-switching observing mode. The primary beam of the IRAM-30m is $\sim$\,27\arcsec~at $\sim$\,90~GHz. The reference position was set to an offset of $\Delta\alpha=351\farcs5$, $\Delta\delta=0\farcs0$~from \object{L1489 IRS}. Single pointing spectra acquired in frequency switching mode toward the reference position showed no contamination from our targeted lines. \object{EQ~0439+360}, \object{3C~84}, \object{4C~32.14} and \object{EQ~0439+360} were used as calibrators for the observations. The pointing was checked approximately every $\sim$\,1--1.5~h on a nearby continuum source. The focus was assessed every $\sim$\,2--2.5~h using a strong source.

The data reduction was conducted using the \texttt{GILDAS CLASS} software on the automatically calibrated data provided by the telescope. The polarizations were averaged resulting in an increase of maps' S/N. Fluxes were converted from antenna temperature corrected from the forward efficiency $T_\text{A}^*$ to main beam temperature $T_\text{mb}$ with the command \texttt{modify~beam\_eff~/ruze} which uses Ruze's equation \citep{ruze1952} to compute beam efficiencies.
We extracted a 30~MHz spectral window around the rest frequency of the line, and removed a 1\textsuperscript{st} order baseline (estimated by eye on the emission filtered from the line) for each spectrum. This baseline-subtracted dataset was then used to produce the maps with the \texttt{CLASS} commands \texttt{table} and \texttt{xy\_map}.

\subsection{Combination of IRAM-30m and NOEMA data}

Single dish and interferometric data were combined using the \texttt{uv\_short} task within \texttt{MAPPING} with a single dish weighting factor of 1. The method employed by this task computes pseudo-visibilities for IRAM-30m data and adds them to NOEMA \textit{uv} tables. It is robust as it enforces the dirty image to have a total positive flux and facilitates the simultaneous deconvolution of both short and long baselines \citep{iram-memo-2008-2}. We note that another commonly used method is the hybridization technique introduced by \cite{weiss2001} which combines interferometric data that are already deconvolved to single dish data. But in this case, the deconvolution lacks crucial information, such as the total flux to be recovered (i.e., the zero-spacing visibility), which is not desirable.

The combined \textit{uv} tables were deconvolved without masks using the Multi Resolution Clean (MRC) algorithm \citep{mrc1988} to CLEAN both compact and extended emission. The gain was set to 0.05, the smoothing ratio to 2, and the stopping criterion to three times the noise level expected in the continuum-free images. All observations presented here are the result of combined data, except for the 3.2~mm continuum presented in Fig.~\ref{fig:continuum}, and the \ce{H^13CN}~($1-0$) and SO~($2_2-1_1$) lines which used NOEMA data only.

\section{Results}
\label{sec:results}

\subsection{Dust continuum}
\label{sec:continuum}
\begin{figure}
    \centering
    \includegraphics[width=\linewidth]{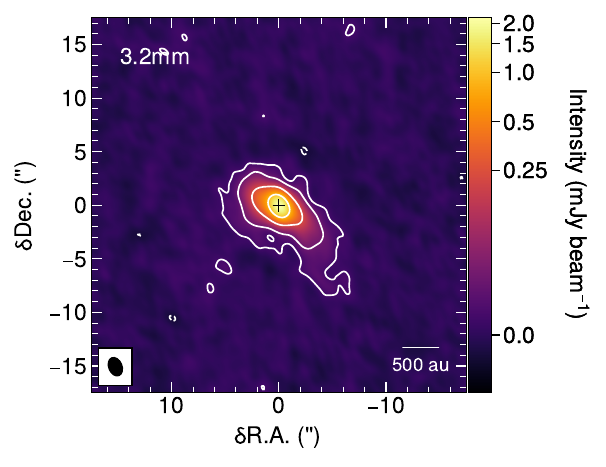}
    \caption{3.2~mm continuum image of \object{L1489 IRS}.  
    The contour levels are [$-$3, 3, 10, 50, 200]$\sigma$ where $\sigma=4.52\,\mu$Jy~beam$^{-1}$. The position (0\arcsec, 0\arcsec) corresponds to the continuum peak localized in $\alpha$(J2000)~=~04\textsuperscript{h}04\textsuperscript{m}43\fs085, $\delta$(J2000)~=~26\degr18\arcmin56\farcs206. The color scale is stretched by the arcsinh function to make faint extended features more visible. The synthesized beam ($1.79\arcsec\times1.24\arcsec$, 22.6\degr) is displayed in the lower left corner while the scale bar in the bottom right corner indicates 500~au.}
    \label{fig:continuum}
\end{figure}
The NOEMA 3.2~mm dust continuum image is presented in Fig.~\ref{fig:continuum}. An elongated disk-like structure is oriented in the northeast southwest direction, as identified in previous 1mm observations \citep{yen2014,sai2020,vanthoff2020,tychoniec2021,ohashi2022,yamato2023}. 
After a correction by the primary beam, and the truncation of the image to 90\% of the beam response (which is enough to contain all emission), we derived the integrated intensity to be $4.8\pm0.5$~mJy and the peak intensity $2.2\pm0.2$ mJy~beam$^{-1}$. The integrated intensity seems consistent with pure thermal emission from dust when extrapolating the SED to these wavelengths \citep{furlan2008,sheehan2017}.

\subsection{Molecular line observations}

\begin{figure*}
    \centering
    \includegraphics[width=0.95\textwidth]{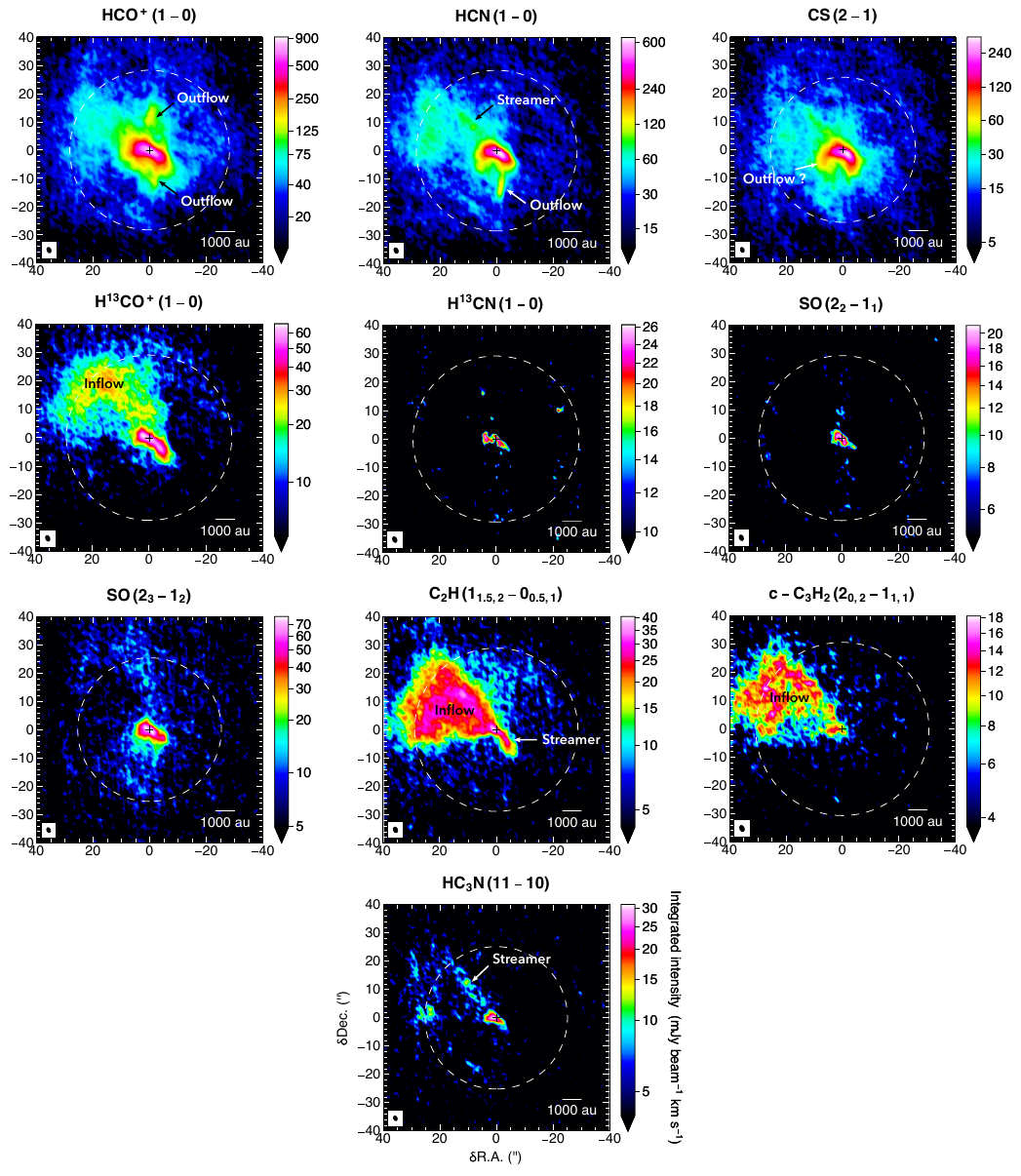}
    \caption{Integrated intensity maps of the primary targeted lines summarized in Table~\ref{table:lines+intensities}. The color scale is stretched by the arcsinh function to make faint extended features more visible. Its minimum is set to $3\sigma$ emission. The synthesized beam of each line is displayed in the lower left corner of each panel, the primary beam with white dashed circles, and the scale bar on the bottom right corner indicates 1\,000~au.}
    \label{fig:mom0-maps}
\end{figure*}

In this study, we focus on the following main detected molecules of our dataset: 
\ce{HCO+}, \ce{H^13CO+}, \ce{HCN}, \ce{H^13CN}, \ce{CS}, \ce{SO}, \ce{C2H}, \ce{HC3N} and \ce{c-C3H2}. These molecules serve as distinct diagnostics in tracing different physical conditions within the observed region. \ce{HCO+}, \ce{H^13CO+}, \ce{HCN}, \ce{H^13CN} and \ce{CS} are well established tracers of cold, dense gas, which are particularly useful for identifying the envelope and potential outflows. \ce{SO} is known to be a shock tracer \citep[e.g.,][and references therein]{garufi2022}, as well as tracing fresh material \citep[e.g.,][and references therein]{haccar2011}, \ce{C2H} and \ce{c-C3H2} trace cavity walls \citep{tychoniec2021}, and \ce{HC3N} is characteristic of infalling material such as "accretion streamers" \citep{pineda2020}. The properties of the synthesized beam, the per-channel rms and the integrated intensities for these lines are provided in Table~\ref{table:lines+intensities}. Integrated intensity maps for these lines are presented in Fig.~\ref{fig:mom0-maps}, while their channel maps are shown in Appendix~\ref{appendix:channel-maps}. To estimate the total integrated intensity of each line, we used the 0\textsuperscript{th} moment map (produced with the new \texttt{GILDAS CUBE} software) of the primary beam-corrected cubes. The velocity ranges used to produce these maps are also detailed in Table~\ref{table:lines+intensities}. The cubes were truncated at a radius corresponding to 20\% of the beam response ($\sim$\,44\farcs3) for most lines, except for the \ce{SO}~($2_2-1_1$) and the \ce{H^13CN}~($1-0$) where a 90\% ($\sim$\,11\farcs6) truncation was applied due to their compact emission. Beyond this 20\% threshold, noise from the edges significantly contaminates the map. Given that the emission is more extended than the primary beam for all lines (except for \ce{SO}~($2_2-1_1$) and \ce{H^13CN}~($1-0$)), we provide a lower limit for the measured flux.

\subsection{Traced structures}

\begin{figure*}
    \centering
    \sidecaption
    \includegraphics[width=12cm]{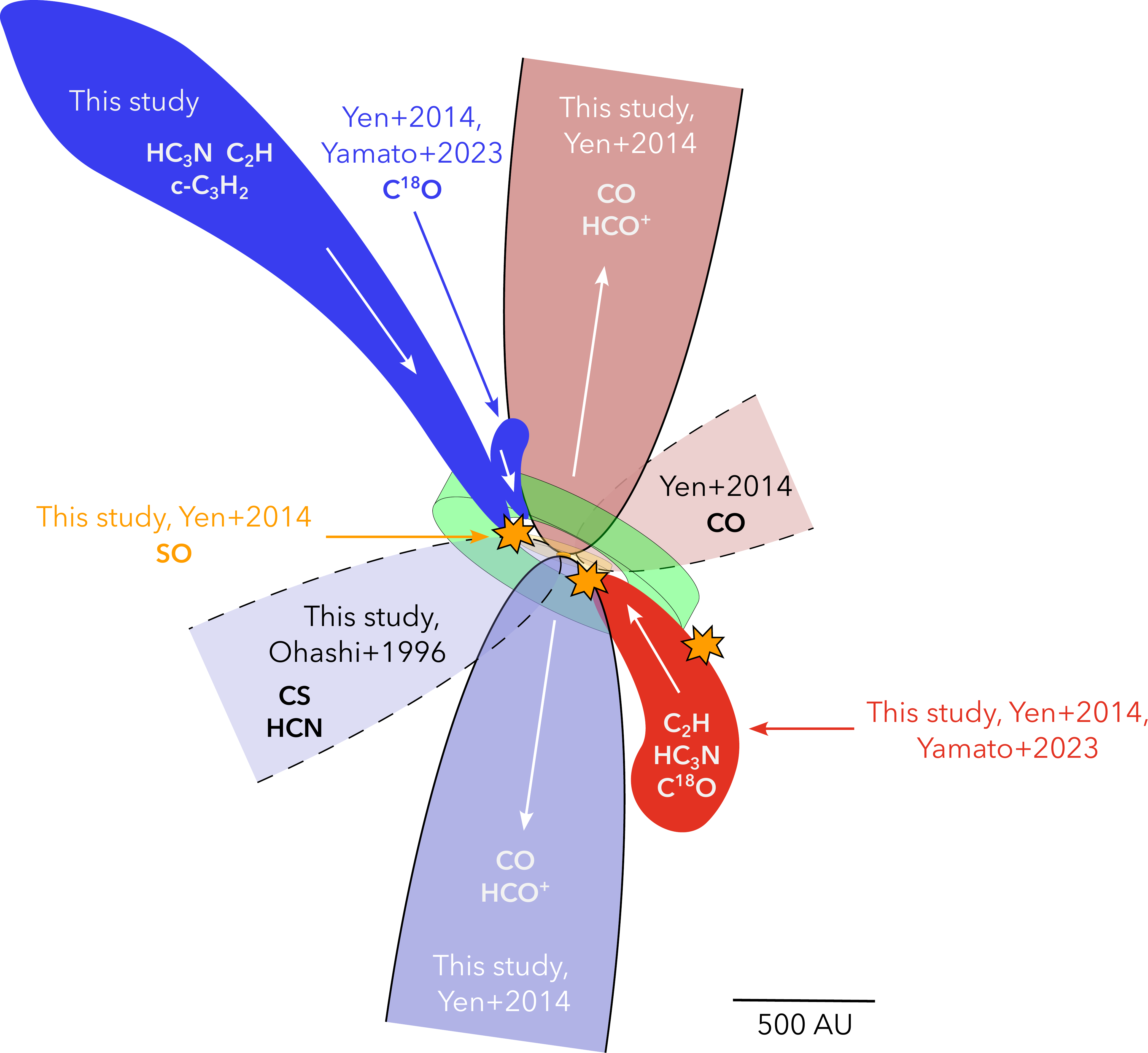}
    \caption{Simplified schematic view of the \object{L1489 IRS} system at scale. The external warped disk (discussed in Sect.~\ref{sec:warp-origin}) is drawn in green, the intermediate disk in yellow and the inner disk in orange. Blue (respectively red) colored structures correspond to blue-shifted (respectively red-shifted) emissions. White arrows indicate the direction of moving material (for the outflow, see Sect.~\ref{sec:outflow}, for the streamers, see Sect.~\ref{sec:secondary-axis} and \ref{sec:analysis}). The accretion shocks (see Sect.~\ref{sec:discussion-so}) are represented in orange. The potential second outflow along the primary axis (see Sect.~\ref{sec:discussion-primary-axis}) is drawn with dashed contours.}
    \label{fig:schematic}
\end{figure*}

\subsubsection{Envelope}
Among the variety of molecular tracers presented in Fig.~\ref{fig:mom0-maps}, we visually distinguish three main groups of traced structures that may reflect the critical density and/or optical depth of each molecular line. Optical depth effects can be appreciated by comparing the emission of \ce{HCO+} and \ce{HCN} main and rare isotopologues. This variety of tracers thus enables us to trace different shells of the protostellar envelope. Firstly, \ce{H^13CN} and \ce{SO} primarily trace relatively compact structures ($\sim$\,1\,000~au) near the disk that could be identified as the innermost envelope. Next, \ce{HC3N}, \ce{c-C3H2}, \ce{C2H} and \ce{H^13CO+} are indicative of substructures within the broader envelope ($\sim$\,5\,000~au), suggesting their association with infalling material (as further detailed in Sect.~\ref{sec:secondary-axis}). Lastly, \ce{HCO+}, \ce{HCN} and \ce{CS} trace the extent of the larger envelope ($\sim$\,10\,000~au).
Additionally, it worth noting that \ce{HCO+} and \ce{HCN} also trace the outflow within the envelope, due to its broader velocity range, which exceeds that of the envelope itself.
Some other structures are also revealed by the targeted molecules, albeit to a lesser extent. For instance, an elongated structure of infalling material, that is an accretion streamer, is apparent in the HCN, \ce{H^13CO+} and CS emission. However, it is rapidly drown in the larger envelope emission, challenging its characterization without contamination. This seems to be avoided using a less abundant tracer like \ce{HC3N}, which was previously identified as a streamer tracer \citep[see][and Sect.~\ref{sec:secondary-axis} for a further description]{pineda2020}.
Additionally, \ce{HCO+}, \ce{HCN}, \ce{CS} and \ce{H^13CO+} exhibit an intermediate bean-shaped envelope as illustrated in Fig.~\ref{fig:mom0-maps}. 
A summary of the different physical components traced by each molecule can be found in Table~\ref{tab:recap-traced-structures}, and illustrated in Fig.~\ref{fig:schematic}. 

\begin{table*}[!ht]
    \centering
    \caption{Main physical components traced by the primary targeted lines of the survey.}
    \label{tab:recap-traced-structures}
    \begin{tabular}{lcccccccc}
        \hline
        \hline
        Line                                & Inner        & Intermediate          & Outer          & Inflows  & Streamers  & Outflow  & Knots & Bubbles \\
                                            & envelope        & envelope              & envelope & & & & \\
        \hline
        \ce{HCO+}~($1-0$)                    & ---            & \yes                  & \yes           & \yes     & \Lsim     & \yes     & \yes   & \yes  \\
        \ce{H^13CO+}~($1-0$)                 & ---            & \yes                  & \yes           & \yes     & \Lsim     & ---      & ---    & \Lsim \\
        \ce{HCN}~($1-0$)                     & ---            & \yes                  & \yes           & \Lsim    & \Lsim     & \yes     & ---    & \yes  \\
        \ce{H^13CN}~($1-0$)                  & \yes           & ---                   & ---            & ---      & ---       & ---      & ---    & ---   \\
        \ce{HC3N}~($11-10$)                  & \yes           & ---                   & ---            & \yes     & \yes      & ---      & ---    & ---   \\
        \ce{C2H}~($1_{1.5,2}-0_{0.5,1}$)         & ---            & ---                   & ---            & \yes     & \yes      & ---      & ---    & ---   \\
        \ce{c}-\ce{C3H2}~($2_{0,2}-1_{1,1}$) & ---            & ---                   & ---            & \yes     & \yes      & ---      & ---    & ---   \\
        \ce{CS}~($2-1$)                      & ---            & \yes                  & \yes           & \Lsim    & \Lsim     & \Lsim    & ---    & \yes  \\
        \ce{SO}~($2_3-1_2$)                  & \yes           & ---                   & ---            & \yes     & ---       & ---      & \yes   & ---   \\
        \ce{SO}~($2_2-1_1$)                  & \yes           & ---                   & ---            & ---      & ---       & ---      & ---    & ---   \\
        \hline
    \end{tabular}
\tablefoot{We distinguish streamers, i.e., elongated structures of infalling material, from inflows, that are wider structures of infalling material. If a structure is barely seen with a molecular tracer, a "\Lsim" symbol is indicated.}
\end{table*}

\subsubsection{Bubbles}
\begin{figure*}
    \centering

    \hspace*{\fill}
    \includegraphics[height=5cm]{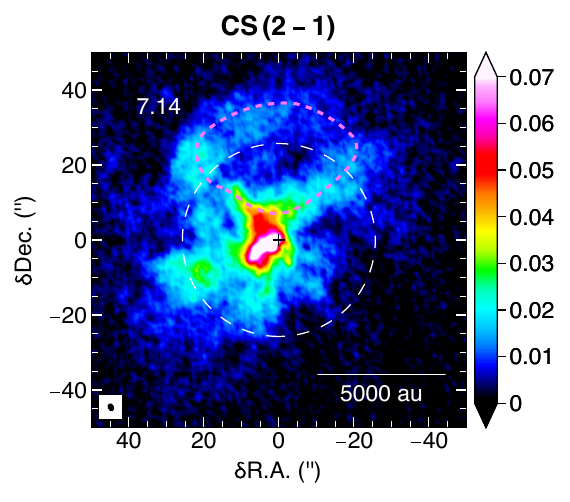}
    \hspace*{\fill}
    \includegraphics[height=5cm]{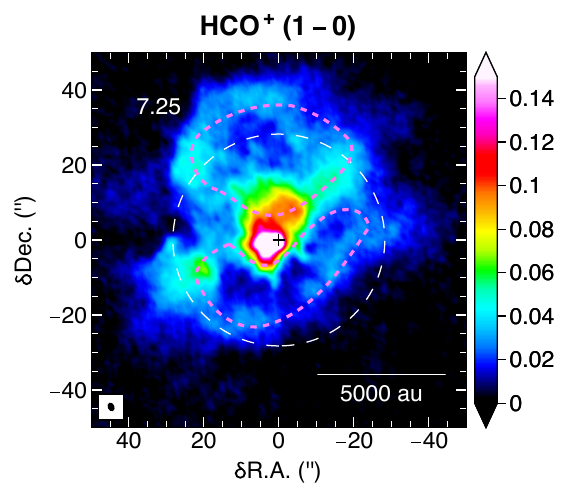}
    \hspace*{\fill}
    \includegraphics[height=5cm]{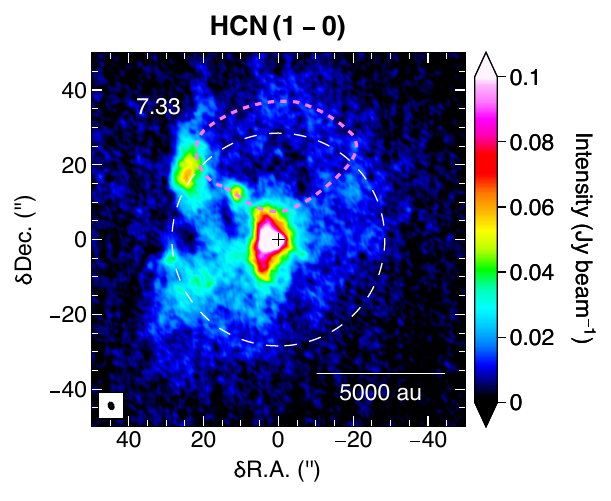}
    \hspace*{\fill}
    
    \caption{Channel maps 
    of \ce{CS}~($2-1$) at 7.14~km~s$^{-1}$ (left), \ce{HCO^+}~($1-0$) at 7.25~km~s$^{-1}$ (middle) and \ce{HCN}~($1-0$) at 7.33~km~s$^{-1}$ (right). The bubbles are indicated with pink dashed lines. The half power primary beam of each line is drawn as the white dashed circle, the synthesized beam is displayed in the lower left corner of each panel and the scale bar in the bottom right corner indicates 5\,000~au. }
    \label{fig:bubbles}
\end{figure*}

Within the channel maps of \ce{CS}, \ce{HCO+} and \ce{HCN}, a distinctive pattern emerges,
as outlined 
in Fig.~\ref{fig:bubbles}. These structures, which we hereafter refer to as "bubbles," consist of two distinct arcs located to the north and south of the targeted source. These bubble features are particularly prominent in the velocity range from 6.8 to 7.7~km~s$^{-1}$ for all three molecules (see Fig.~\ref{fig:bit-cs}, \ref{fig:bit-hcop}, and \ref{fig:bit-hcn}). 
These bubble structures are characterized by their substantial width, extending to $\sim$\,10\,000~au, that is beyond the half-power primary beam. It should be noted that these structures are not clearly visible in IRAM-30m data (maybe due to the dilution of these thin $\sim 10\arcsec$ structures in the $\sim$\,30\arcsec half power 30m beam width), thus mosaic observations with NOEMA are required to confirm their presence. These unique structures are noteworthy features observed on both sides of the targeted low-mass YSO, and their properties and implications warrant further investigation as further described in Sect.~\ref{sec:bubbles-discussion}.

\begin{figure}[!t]
    \centering
     \includegraphics[width=\linewidth]{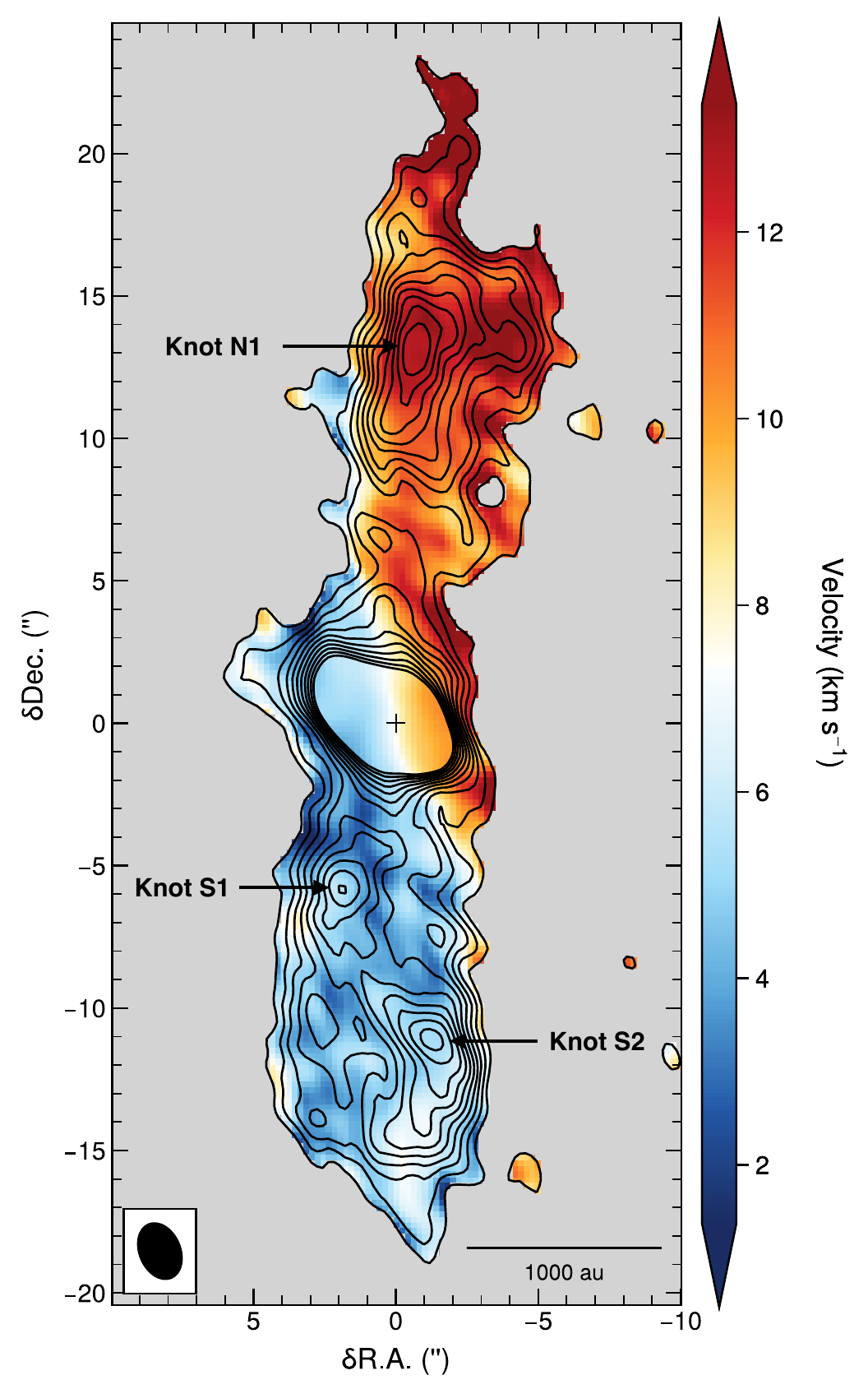}
    \caption{Moment 0 map (contours) overlaid on the moment 1 map (color) for the outflow seen in the \ce{HCO+} $J=1-0$ emission 
    The contour levels are 5$\sigma$ to 25$\sigma$ by 2$\sigma$ steps where $\sigma=3.22$ mJy~beam$^{-1}$~km~s$^{-1}$. The central velocity is set to the $v_\text{LSR}=7.37$~km~s$^{-1}$. The synthesized beam is displayed in the lower left corner. The scale bar on the bottom right corner indicates 1\,000~au.}
    \label{fig:hcop10_outflow}
\end{figure}

\subsubsection{Outflow}
\label{sec:outflow}
Figure~\ref{fig:hcop10_outflow} displays the integrated intensity map (i.e., moment 0 map) of \ce{HCO+}~$(1-0)$ in contours, and its velocity map (i.e., moment 1 map) in background, over the velocity range specified in Table~\ref{table:lines+intensities} excluding the envelope component between 5.7 and 9.3~km~s$^{-1}$. An outflow clearly stands out with a P.A. of $-8\degr$, which is perpendicular to the inner disk major axis \citep[P.A.~$=82\degr$,][]{yamato2023}. This outflow was previously identified with \ce{^12CO}~$(2-1)$ emission but with a less extended emission than the \ce{HCO+}~$(1-0)$ presented here \citep{yen2014}. Specifically, the \ce{^12CO}~$(2-1)$ seems to trace the walls of the cavity while \ce{HCO+} traces the material within it, in which a knot on the northern part (referred to as "knot N1" in Fig.~\ref{fig:hcop10_outflow}) and two knots on the southern part are detected (designated as "knots S1 and S2" in Fig.~\ref{fig:hcop10_outflow}). Interestingly, "knot S2" appears to be consistent with the Herbig-Haro object \object{HH 360A}, which has been previously detected in \ce{H2} and S[II] \citep[see][and Fig.~\ref{fig:herschel-HH}]{gomez1997}. Additionally, the velocity gradient at the bottom of the blue lobe (about 2~km~s$^{-1}$) indicates a rotation of the cavity. The SO~($2_3-1_2$) emission also reveals two knots at $\delta$Dec$=\pm22\arcsec$ that are slightly more distant than those in \ce{HCO^+}. The outflow is also traced by HCN, but the hyperfine components are blended in the spectra due to the wings of the outflow, making a detailed dynamical study challenging. We note that the second branch of the V-shape (P.A.~$=-60$\degr) in the northern \ce{^12CO}~$(2-1)$ emission \citep[see Fig.~\ref{fig:schematic} in this work and Fig.~2 in][]{yen2014} is not traced with the molecules we targeted here. This emission may be hidden in the envelope if its velocity range is not broad enough.

\subsubsection{Quadrupolar flow}
\begin{figure}
    \centering
    \includegraphics[width=\linewidth]{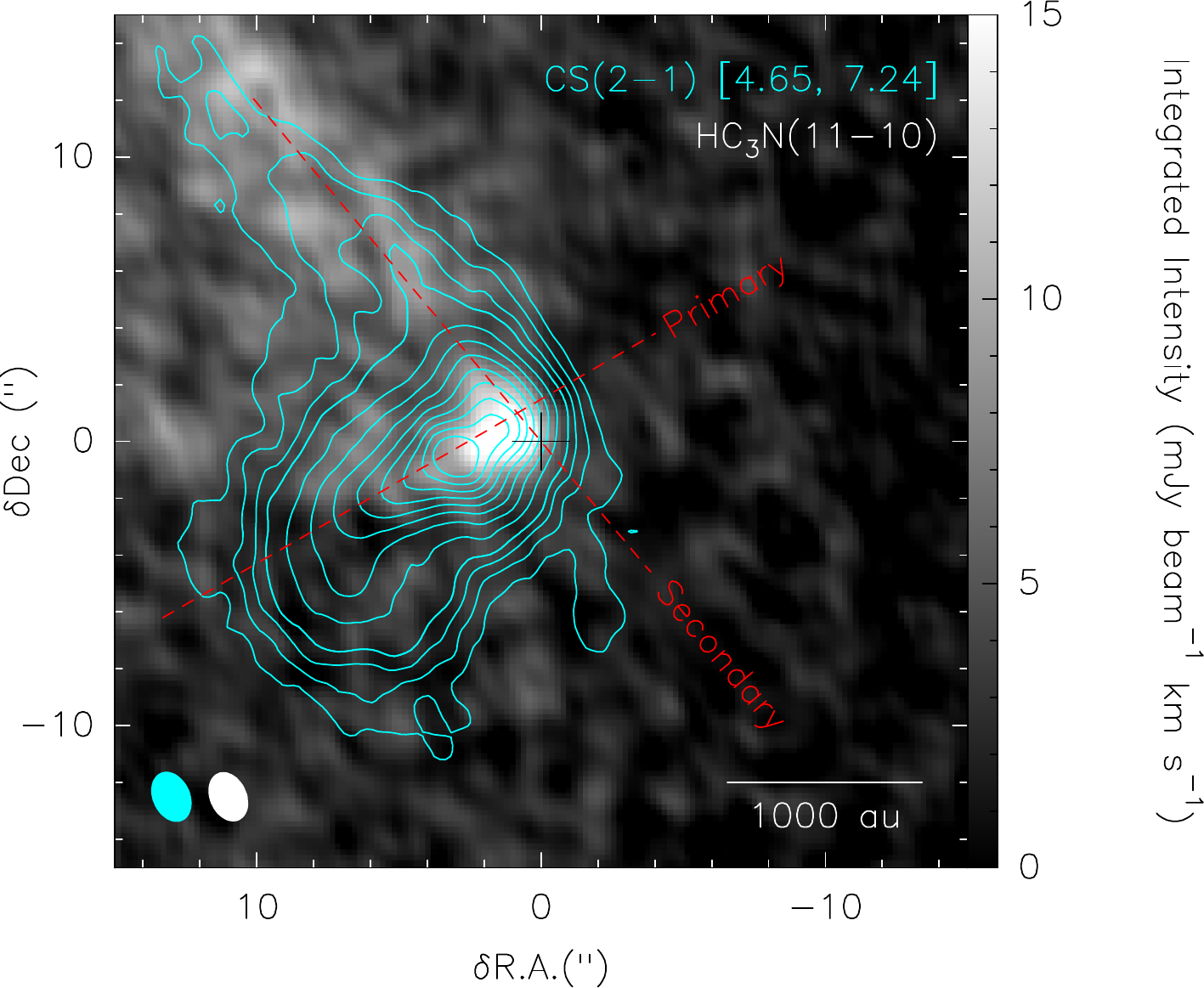}
    \caption{\ce{CS(2-1)} blue-shifted ($4.65-7.24$~km~s$^{-1}$) integrated intensity map (contours) overlaid on the \ce{HC3N(11-10)} integrated intensity map (background, colorbar saturated between 0 and 15~mJy~beam$^{-1}$~km~s$^{-1}$). The contour levels are 30$\sigma$ to 45$\sigma$ by 5$\sigma$ steps then 45$\sigma$ to 125$\sigma$ by 10$\sigma$ steps where $\sigma=1.0$~mJy~beam$^{-1}$~km~s$^{-1}$. The two axis of the quadrupolar flow are denoted with dashed red lines. The synthesized beams are displayed in the lower left corner. The scale bar on the bottom right corner indicates 1\,000~au.}
    \label{fig:quadrupolar-flow}
\end{figure}

\begin{figure*}
    \centering
    \sidecaption
    \hspace*{\fill}
    \includegraphics[width=6cm]{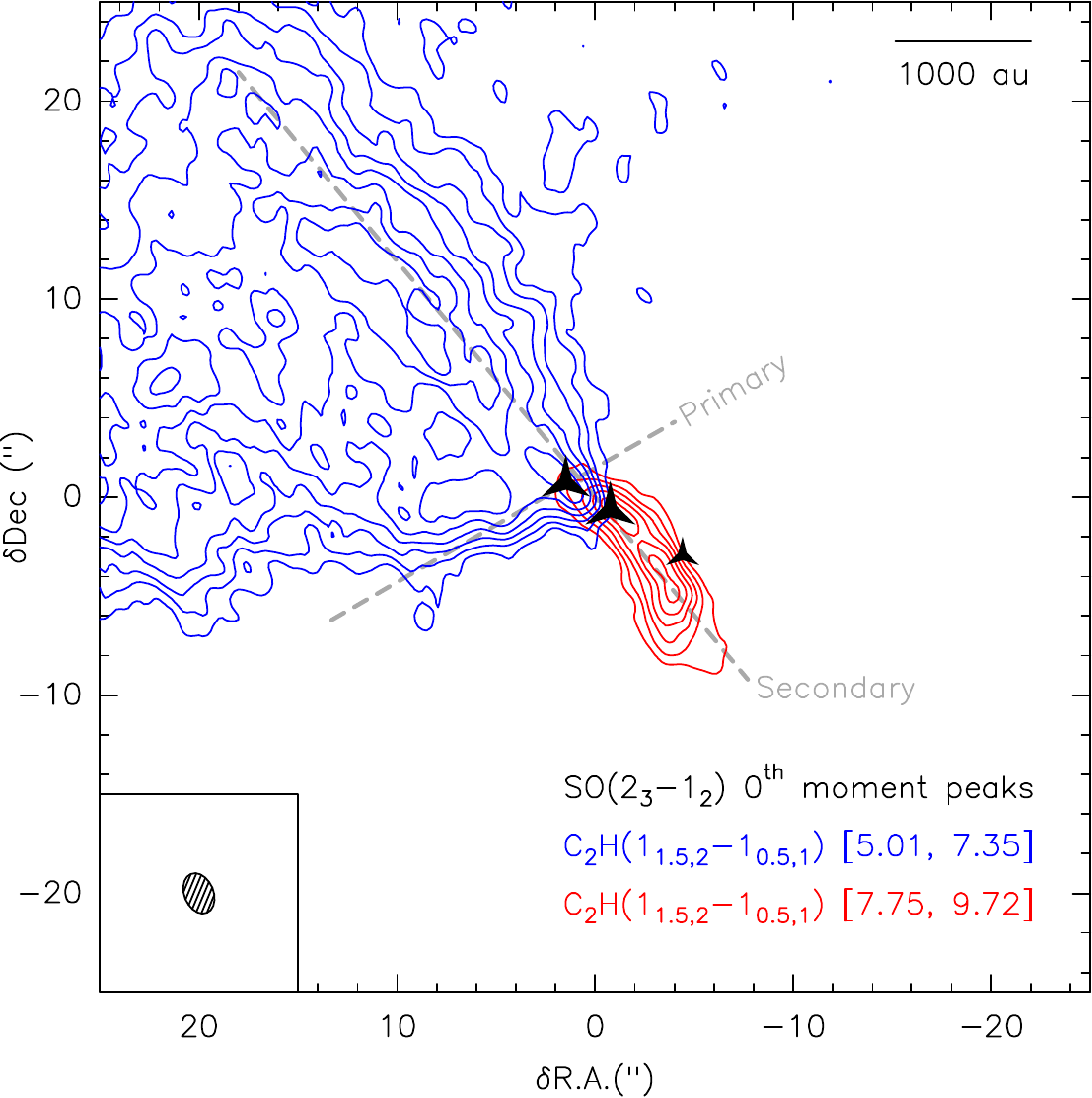}
    \hspace*{\fill}
    \includegraphics[width=6cm]{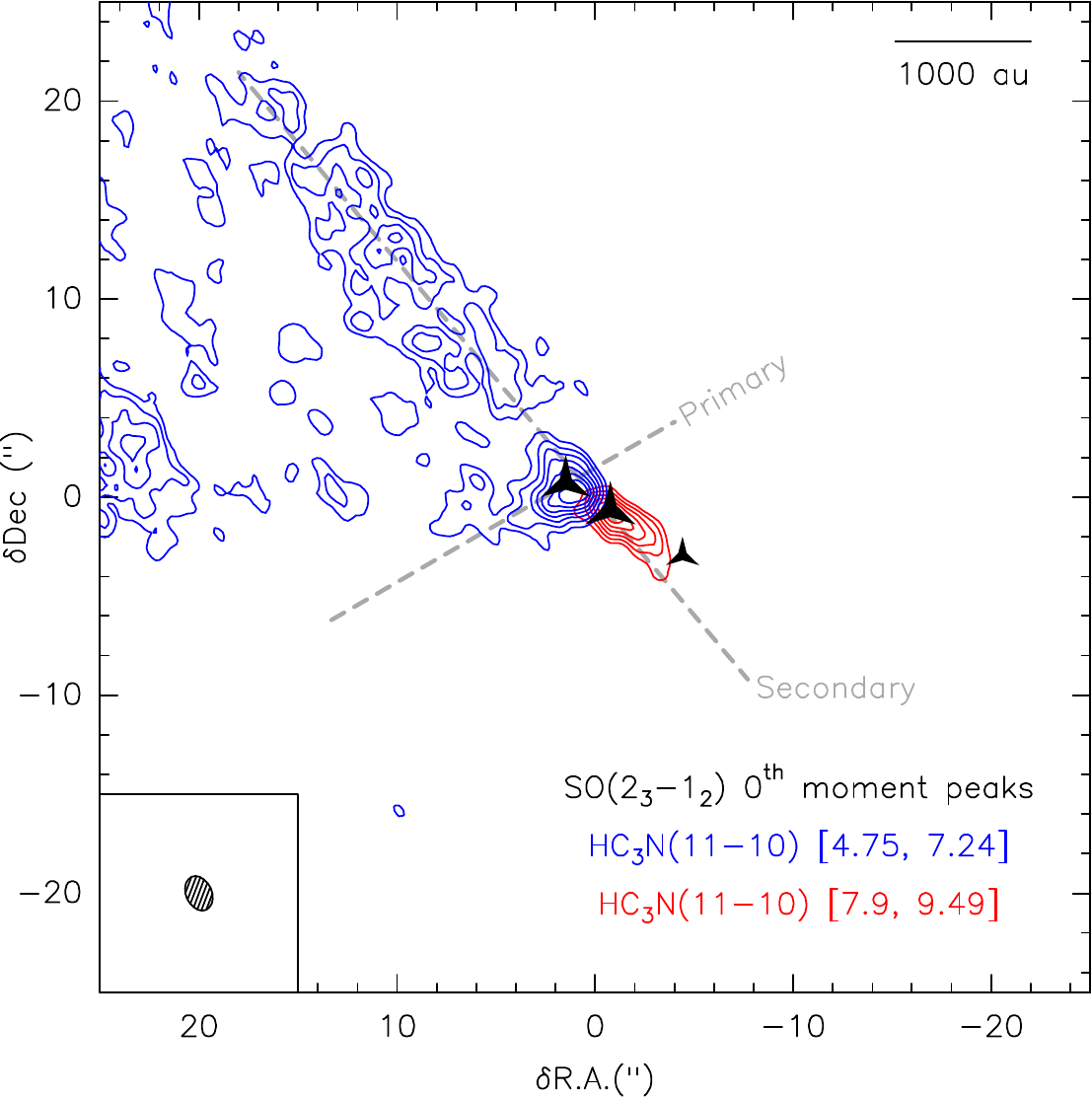}
    \hspace*{\fill}
    \caption{Integrated intensity contours maps 
    of the \ce{C2H}~($1_{1.5,2}-0_{0.5,1}$) emission, on the left, and the \ce{HC3N}~($11-10$) emission, on the right. The beam of each line is displayed in the lower left corner of each panel. The three brightest peaks of the \ce{SO}~($2_3-1_2$) integrated map are shown by black triangles, whose size depends on their intensity relative to the overall maximum. The primary and the secondary axis are represented by the grey dashed-lines.
    \textbf{Left:} \ce{C2H}~($1_{1.5,2}-0_{0.5,1}$) blue-shifted emission ($5.01-7.35$ km~s$^{-1}$) in blue contours and red-shifted emission ($7.75-9.72$ km~s$^{-1}$) in red contours. The levels are 10$\sigma$ to 33$\sigma$ by 3$\sigma$ steps where $\sigma=0.97$ mJy beam$^{-1}$ km~s$^{-1}$. \textbf{Right:} \ce{HC3N}~($11-10$) blue-shifted emission ($4.75-7.24$ km~s$^{-1}$) in blue contours and red-shifted emission ($7.77-9.49$ km~s$^{-1}$) in red contours. The contour levels are 7$\sigma$ to 19$\sigma$ by 2$\sigma$ steps where $\sigma=0.89$ mJy beam$^{-1}$ km~s$^{-1}$.\\}
    \label{fig:streamer-shocks}
\end{figure*}

A quadrupolar flow-like structure was identified in previous \ce{H2}~$(1-0)$~S(1) observations at 2.12 $\mu$m obtained with the United Kingdom Infrared Telescope (UKIRT), where two major axis were characterized as "primary" and "secondary" \citep{lucas2000}. We also detect this structure in our data, as shown in Fig.~\ref{fig:quadrupolar-flow}, and further described below. Subsequently, we consistently adopt the same nomenclature for these axes.

\paragraph{Primary axis:}
\label{sec:primary-axis}
Elongated emission toward the southeast along a P.A.~$=-60\degr$ is prominently observed in the blue-shifted emission of \ce{CS} \citep[which was also observed in the same emission line with the Nobeyama Millimeter Array (NMA), by][with 8\arcsec~of resolution]{ohashi1996}, \ce{HCN} and \ce{SO}~$(2_3-1_2)$ (see Fig.~\ref{fig:quadrupolar-flow}, \ref{fig:bit-cs}, \ref{fig:bit-hcn}, \ref{fig:bit-so2312}) with a projected length of $\sim$\,2\,100~au. Its extent is consistent with the \ce{H2} emission pattern reported by \citet{lucas2000}. In particular, the emission peak in the integrated blue-shifted CS map is spatially distinct from the disk's center (located at (0\arcsec, 0\arcsec)) and is instead concentrated at the southeastern edge of the 3.2~mm continuum shown in Fig.~\ref{fig:continuum}. Furthermore, the \ce{HC3N} emission overlaps with the \ce{CS} along the primary axis between the emission peak and the secondary axis, and are thus collocated within the continuum emission.

\paragraph{Secondary axis:}
\label{sec:secondary-axis}
The integrated map of \ce{C2H} is shown on the left panel of Fig.~\ref{fig:streamer-shocks}. The red-shifted component shows an elongated structure with a projected length of $\sim$\,1\,900~au oriented toward the southwest of the protostar(s). It matches the red-shifted component of the \ce{C^18O}~$(2-1)$ emission reported by \cite{yen2014}, which they characterized as infalling material. The right panel of  Fig.~\ref{fig:streamer-shocks} shows the integrated map of \ce{HC3N}. Its red-shifted component is along the \ce{C2H} and the \ce{C^18O}~$(2-1)$ emission, but is more than twice smaller than \ce{C2H} (projected length of $\sim$\,800~au). The SO channel maps (see Fig.~\ref{fig:bit-so2312} and Fig.~\ref{fig:bit-so2211}) between 8 and 8.6~km~s$^{-1}$ reveal a filamentary structure within the \ce{C2H} emission that connects the disk to a bright blob located at a projected distance of 785~au from the disk. A trace of this blob is also seen on the \ce{c-C3H2} channel map at 7.98~km~s$^{-1}$ (with a 6$\sigma$ emission peak in $\delta$RA$=-4.10\arcsec$, $\delta$Dec.$=-2.30\arcsec$, see Fig.~\ref{fig:bit-cc3h2}). 
Another filamentary structure on top of the SO filament is seen in HCN for the same velocity range but the blob does not appear in the HCN emission (see Fig.~\ref{fig:bit-hcn}).\\
Regarding the blue-shifted component, the \ce{C2H} emission unveils a much wider structure at the northeast of the protostar(s) (projected length of $\sim$\,5\,500~au) than the \ce{C^18O}~$(2-1)$ in \citet{yen2014}. \citet{sai2022} also mapped the \ce{C^18O}~$(2-1)$ emission at larger scales and revealed a much more extended emission than \ce{C2H}. They modeled the kinematics of this wide emission and concluded to slow infalling material at a speed $\sim$\,0.4 of the free-fall velocity. The blue-shifted emission of \ce{HC3N} is as extended as \ce{C2H} but traces a V-shape structure with a smaller opening angle than the \ce{C2H} emission. The northern part of the V-shape, elongated along the secondary axis (P.A.$=40\degr$), especially stands out on the channel map of \ce{HC3N} (between 7.05 and 7.23 km~s$^{-1}$, see Fig.~\ref{fig:bit-hc3n}) and is also seen on the channel map of \ce{C2H} (between 7.13 and 7.35 km~s$^{-1}$, see Fig.~\ref{fig:bit-c2h}) and \ce{c-C3H2} (between 7.07 and 7.29 km~s$^{-1}$, see Fig.~\ref{fig:bit-cc3h2}). This emission is reminiscent of a streamer, likely nested within the broader, extended envelope.
Indeed, this thin structure is also visible on the blue-shifted channels maps of \ce{H^13CO^+}, HCN and CS (see Fig.~\ref{fig:bit-h13cop}, \ref{fig:bit-hcn}, \ref{fig:bit-cs} and \ref{fig:quadrupolar-flow}) but is rapidly drown in the envelope emission as the velocity increases. The southern part of the V-shape, elongated along a P.A.$=84\degr$, shows a clump around (24\arcsec,~2\arcsec), corresponding to velocity channels between 6.67 and 6.86~km~s$^{-1}$ (see Fig.~\ref{fig:bit-hc3n}). This velocity range corresponds to the $v_\text{LSR}$ of the parental cloud \citep[$6.6-6.8$~km~s$^{-1}$,][]{wu2019}.

Finally, we note a spatial correlation between the two brightest peaks of emission of the integrated intensity map of SO~($2_3-1_2$) displayed with black triangles in Fig.~\ref{fig:streamer-shocks}, and the base of the blue and red-shifted emission of both \ce{HC3N} and \ce{C2H}.

\section{Analysis}
\label{sec:analysis}
\subsection{Position-velocity diagrams}
\begin{figure*}
    \centering
    \includegraphics[width=\textwidth]{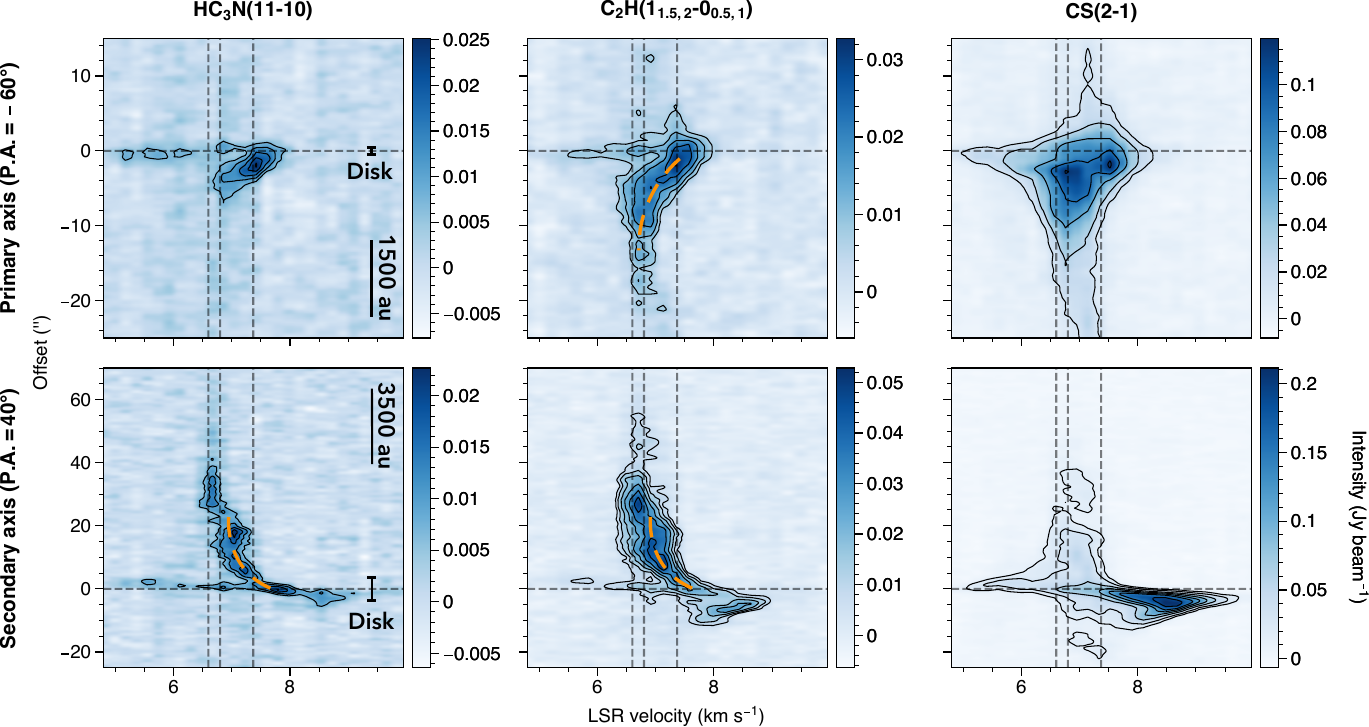}
    
    \caption{Position-velocity (PV) diagrams along the primary axis (first row) and the secondary axis (second row) of the quadrupolar flow. Arm-shape structures are drawn with dashed orange lines. The disk spatial extension is given by the black vertical line. The vertical dotted lines correspond to the core $v_\text{LSR}=6.6-6.8$~km~s$^{-1}$ \citep{wu2019} and the disk $v_\text{LSR}=7.37$~km~s$^{-1}$, the horizontal to the 0\arcsec~offset. Offsets are positive toward the north. \textbf{First column:} \ce{HC3N} emission. The contour levels are [5, 8, 11, 14, 17]$\sigma~$ where $\sigma=1.45$ mJy~beam$^{-1}$. \textbf{Second column:} \ce{C2H} emission. The contour levels are [5, 8, 12, 17, 23, 30]$\sigma~$ where $\sigma=1.39$ mJy~beam$^{-1}$. \textbf{Third column:} \ce{CS} emission. The contour levels are [10, 20, 40, 60, 80, 100, 130]$\sigma~$ where $\sigma=1.44$ mJy~beam$^{-1}$.}
    \label{fig:quadrupolar-pv}
\end{figure*}

We extracted the position-velocity (PV) diagrams along the two major axis of the quadrupolar flow, corresponding to P.A. of $-60$\degr~(primary axis) and 40\degr (secondary axis), for \ce{HC3N}, \ce{C2H} and \ce{CS} whose emissions are structured along those two axis. The results are shown in Fig.~\ref{fig:quadrupolar-pv}. On the primary axis, the \ce{HC3N} emission is concentrated between [$-6\arcsec$,~$2\arcsec$] offsets but shows a velocity gradient from the core/cloud $v_\text{LSR}=6.6-6.8$~km~s$^{-1}$ \citep{wu2019} to 8~km~s$^{-1}$. This same structure is seen in the \ce{C2H} emission over the same positional offsets, but connects to a larger emission at $6.6-6.8$~km~s$^{-1}$ extending to $-22\arcsec$, resulting in an arm-shape structure. This feature is not clearly seen in the \ce{CS} emission. For the secondary axis, an arm-shape is also present on the blue-shifted side in \ce{HC3N} and \ce{C2H}, while \ce{CS} shows a prominent emission at red-shifted velocities that is also partly seen in \ce{C2H}, and to a lesser extent in \ce{HC3N}.

Those arms-like structures bear resemblance with infalling material, indicated by the increasing velocity of the material as it approaches the disk. To ascertain the presence of infalling material, we modeled the kinematics using two models, and attempted to reproduce the observations, as further described in the following.

\subsection{Streamline model}
\label{sec:streamline-model}
We used the Newtonian analytic solutions provided by \citet{mendoza2009} to model infalling material within a rotating cloud, converging toward a central object where gravitational forces dominates. 
This model, hereafter referred as the "streamline model," is an extension of the Ulrich profile \citep{ulrich1976,mendoza2004}, a widely employed framework for characterizing material within systems featuring both envelopes and disks \citep[e.g.,][]{yen2017,yen2019,pineda2020,garufi2022,thieme2022,valdivia-mena2022,kido2023,Flores2023}.
We used the implementation of \citet{pineda2020}\footnote{\url{https://github.com/jpinedaf/NOEMA_streamer_analysis}} in the following. The input parameters of the model are:
\begin{itemize}
\item the initial angular velocity of the cloud $\Omega_0$, 
\item the mass of the central object $M_\star$, 
\item the initial position $(r_0,~\theta_0,~\varphi_0)$ and radial velocity $v_{r,0}$ of the infalling mass in spherical coordinates. 
\end{itemize}

$r_0$ is the initial radius with respect to the central object. Initially in the model, $\theta_0$ is the polar angle with respect to the $z$-axis (oriented toward positive Dec.) and $\varphi_0$ is the azimuthal angle with respect to the $x$-axis (oriented toward negative R.A.). The rotational axis of the material is thus the $z$-axis. However, a user-defined rotational axis can be chosen through its inclination $i$ and its P.A.. In that case, the inclination value is rotated around the $x$-axis, followed by a similar rotation of the P.A. around the $y$-axis. This results in definitions of $\theta_0$ and $\varphi_0$ in the disk's reference system \citep[as described in][]{pineda2020}. The model's output is the trajectory and the velocity of the mass along the streamline in cartesian coordinates.

No previous values of the initial angular velocity of the cloud $\Omega_0$ are available in the literature. In the model, $\Omega_0$ is only used for calculating the centrifugal radius $r_u$ using:
\begin{equation}
    \centering
    r_u = \dfrac{r_0^4\,\Omega_0^2}{GM_\star}
    \label{eq:centrifugal-radius}
\end{equation}
In their study, \citet{sai2022} used a centrifugal radius $r_u=600$~au for their disk-and-envelope model, which corresponds to the radius of the external Keplerian disk \citep{sai2020}. As the overall shape of the projected trajectory on the sky does not change much with $r_u$ values of few hundreds of au, we adopted the same here and we inferred $\Omega_0$ from Eq.~\ref{eq:centrifugal-radius}, using $r_u=600$~au. Concerning the central mass $M_\star$, \cite{yamato2023} derived $M_\star=1.7\pm0.2~M_\odot$. We tested three values in this range: $1.5~M_\odot$, $1.7~M_\odot$ and $1.9~M_\odot$. As the central mass increases, the projected trajectory on the sky does not substantially change the shape of the streamline. We thus elect to adopt $M_\star=1.7~M_\odot$. For the radial velocity $v_{r,0}$, we also tested three values: 0 km~s$^{-1}$, $0.4\,v_{ff,\,0}$ \citep[slow infall identified by][]{sai2022} and $v_{ff,\,0}$ where $v_{ff,\,0}$ is the free fall velocity at $r_0$ computed as:
\begin{equation}
    v_{ff,\,0} = \sqrt{\dfrac{\,2GM_\star}{r_0}}
    \label{eq:v-free-fall}
\end{equation}
The increase of the initial radial velocity does not change the overall shape of the streamline (see Fig.~\ref{fig:streamline-compare-vel}).
However, it has a clearer impact on the observed velocity profile along the projected radius 
which can be directly compared to the observations, and thus serve to enhance the precision in the parameter constraints (see Fig.~\ref{fig:streamline-compare-vel}). 
The limitation of reduced data points in using a specific P.A. cut for PV diagrams can be overcome with the Kernel Density Estimation (KDE) technique, providing a global representation of PV diagram variations \citep{pineda2020}. We applied this technique, using the \texttt{scipy} python module \citep{scipy2020}, on the velocity map (moment 1 map) within a user-defined mask corresponding to the region of interest for the modeling.
Finally, we tested three different rotational axis :
\begin{itemize}
    \item the default $z$-axis : $i=0\degr$, P.A.~$=0\degr$, hereafter the $(\mathcal{R}_z)$ axis ;
    \item the outflow-defined axis, assuming its inclination perpendicular to the disk \citep[as in][]{valdivia-mena2022} : $i=18\degr$, P.A.~$=-8\degr$, hereafter the $(\mathcal{R}_o)$ axis ;
    \item the external-disk-defined axis, assuming its inclination and its P.A. perpendicular to the disk : $i=18\degr$, P.A.~$=-36\degr$, hereafter the $(\mathcal{R}_d)$ axis.
\end{itemize}

\subsubsection{Along the primary axis}
\label{sec:streamline-model-CS}

\begin{figure*}[!ht]
    \centering
    \sidecaption
    \includegraphics[width=12cm]{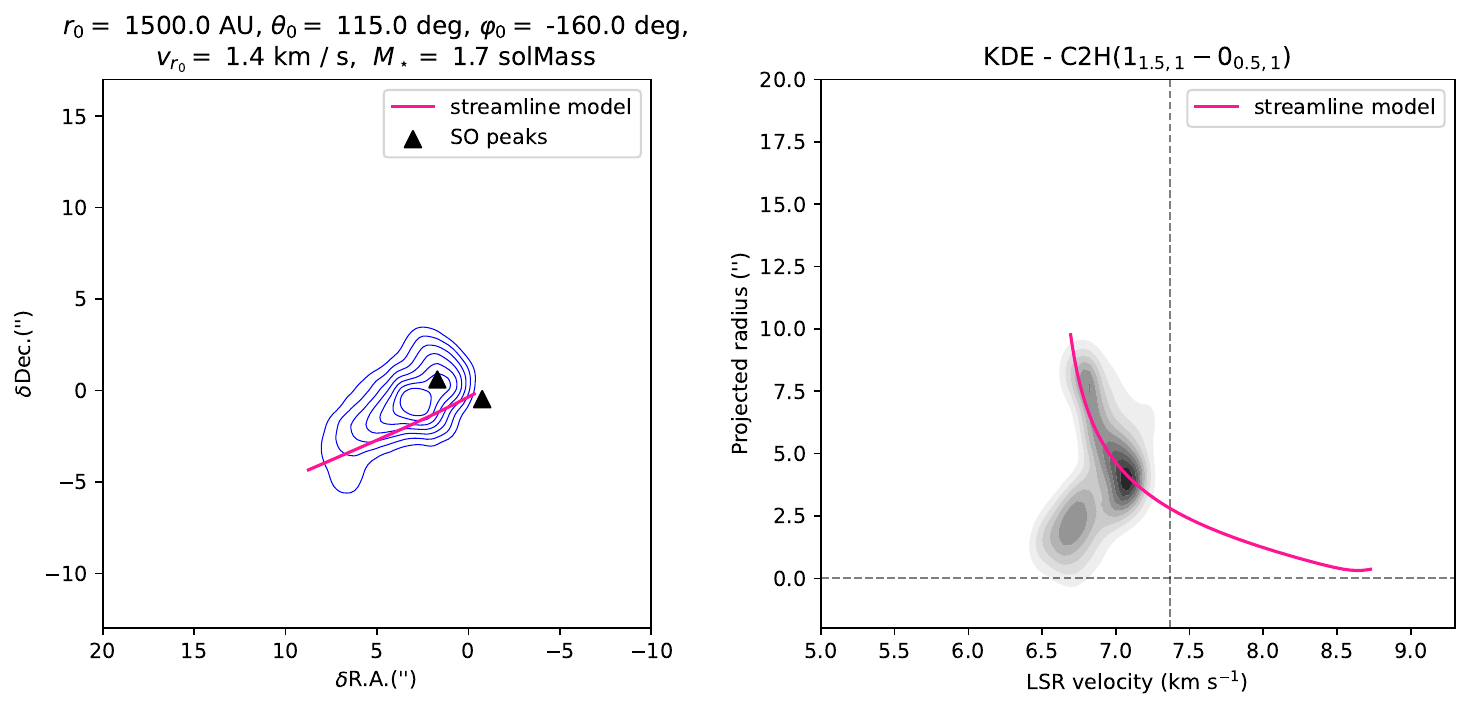}
    \caption{Streamline model of the \ce{CS} blue-shifted emission for the set of parameters $(r_0,~\theta_0,~\varphi_0,~v_{r,0})=$ (1500~au, 115\degr, $-160$\degr, 1.4~km~s$^{-1}$) around the $(\mathcal{R}_z)$ axis. See Sect.~\ref{sec:streamline-model-CS} for more details. \textbf{Left:} Theoretical projected trajectory (in pink) from the streamline model overlayed on the \ce{CS} blue-shifted emission and the two brightest peaks of the SO emission of Fig.~\ref{fig:streamer-shocks}. The contour levels are 65$\sigma$ to 125$\sigma$ by 10$\sigma$ steps where $\sigma=1.0$~mJy~beam$^{-1}$~km~s$^{-1}$. \textbf{Right:} Theoretical line of sight velocity profile from the streamline model overlayed on the KDE of the \ce{C2H} blue-shifted velocity map. The vertical dotted line corresponds to the disk $v_\text{LSR}=7.37$~km~s$^{-1}$, the horizontal to the 0\arcsec~offset.}
    \label{fig:streamline-model-CS}
\end{figure*}

First, we conducted a large exploration focused on refining
the geometry by examining only the projected trajectory on the sky. We compared the output to the blue-shifted \ce{CS} emission, which is the best tracer recovering this elongated emission along this axis. We fixed $r_0=1\,500$~au, as keeping a $\sim10^4$~au long streamline does not alter the overall shape of the projected trajectory in the sky, and $v_{r,0}=0$~km~s$^{-1}$. With the $(\mathcal{R}_z)$ axis, we varied $\theta_0$ between 90\degr~and 180\degr~by 5\degr~increments and $\varphi_0$ between $-90$\degr~and $-270$\degr~by 5\degr~increments. This initial step enabled us to refine the geometric parameters, leading to a subsequent second investigation with finer parameter ranges: $\theta_0$ between 100\degr~and 130\degr~by 1\degr~increments and $\varphi_0$ between $-120$\degr~and $-170$\degr~by 5\degr~increments. $r_0$ was kept fixed at 1500~au.  We then assessed the correspondence between the velocity profile along the projected radius toward the center and the KDE. The latter is computed on the \ce{C2H} blue-shifted ($4.65-7.37$~km~s$^{-1}$) velocity map, within a mask delineated by the first contour level of the blue-shifted CS emission in Fig.~\ref{fig:streamline-model-CS}. The choice of using \ce{C2H} instead of \ce{CS} for the KDE computation was motivated by the PV diagrams (see Fig.~\ref{fig:quadrupolar-pv}). Notably, the arm-like structure is more pronounced in the \ce{C2H} emission than the \ce{CS} emission, possibly indicating that \ce{C2H} traces deeper layers than \ce{CS}, likely due to its lower optical depth (the \ce{C2H}~($1_{1.5, 2}-0_{0.5, 1}$) and \ce{CS}~($2-1$) Einstein coefficients from the CDMS are $1.5\times 10^{-6}$~s$^{-1}$ and $1.7\times 10^{-5}$~s$^{-1}$ respectively). Note this mask implies a contamination by the disk and/or the inner envelope, resulting in the presence of velocities at $\sim6.5$~km/s at small projected radii in Fig.~\ref{fig:streamline-model-CS}. As we searched for a better constraint on the velocity profile, we varied $v_{r,0}$ accross three values: 0~km~s$^{-1}$, $0.4\,v_{ff,\,0}=0.6$~km~s$^{-1}$ and $v_{ff,\,0}=1.4$~km~s$^{-1}$.

We found degenerated results, where multiple parameter sets falling within these ranges match either with the projected trajectory in the plane of sky or the KDE. Figure~\ref{fig:streamline-model-CS} illustrates one of the most visually compelling models which reconciles the two. However, the modeled trajectory is not very satisfying as it does not suit well the observed emission. This discrepancy remains when considering the $(\mathcal{R}_o)$ and $(\mathcal{R}_d)$ axis. Consequently, the infalling material hypothesis along the primary axis remains uncertain.

\subsubsection{Along the secondary axis}
\label{sec:streamline-model-HC3N}

\begin{figure*}[!ht]
    \centering
    \sidecaption
    \includegraphics[width=12cm]{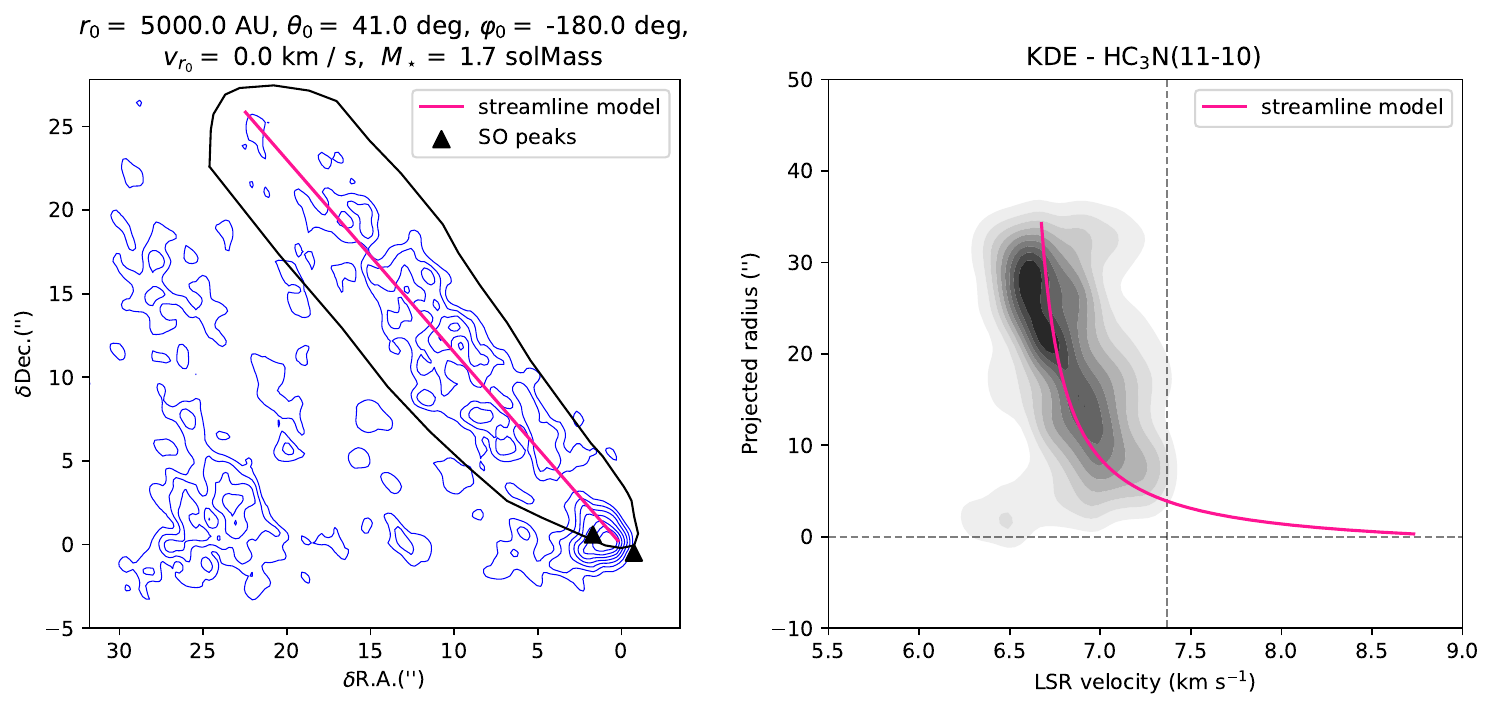}
    \caption{Streamline model of the \ce{HC3N} blue-shifted emission for the set of parameters $(r_0,~\theta_0,~\varphi_0,~v_{r,0})=$ (5000~au, 41\degr, $-180$\degr, 0.0~km~s$^{-1}$) around the $(\mathcal{R}_z)$ axis. See Sect.~\ref{sec:streamline-model-HC3N} for more details. \textbf{Left:} Theoretical projected trajectory (in pink) from the streamline model overlayed on the \ce{HC3N} blue-shifted emission and the two brightest peaks of the SO emission of Fig.~\ref{fig:streamer-shocks}. The contour levels are 7$\sigma$ to 19$\sigma$ by 2$\sigma$ steps where $\sigma=0.89$ mJy beam$^{-1}$ km~s$^{-1}$. The black contour delimits the mask used to compute the KDE. \textbf{Right:} Theoretical line of sight velocity profile from the streamline model overlayed on the KDE of the \ce{HC3N} blue-shifted velocity map. The vertical dotted line corresponds to the disk $v_\text{LSR}=7.37$~km~s$^{-1}$, the horizontal to the 0\arcsec~offset.}
    \label{fig:streamline-model-HC3N}
\end{figure*}

We adopted a similar methodology for the secondary axis, using this time the \ce{HC3N} emission (the best tracer recovering this elongated emission along this axis), by first starting with a wide parameter exploration.
We fixed $r_0=5\,000$~au and $v_{r,0}=0$~km~s$^{-1}$. With the $(\mathcal{R}_z)$ axis, we varied $\theta_0$ between 35\degr~and 75\degr~by 5\degr~increments and $\varphi_0$ between $-90$\degr~and $-180$\degr~by 5\degr~increments. Then we carried out a second exploration with finer ranges on this better constrained geometry: $\theta_0$ between 40\degr~and 50\degr~by 1\degr~increments and $\varphi_0$ between $-135$\degr~and $-180$\degr~by 5\degr~increments. $r_0$ was still fixed to 5\,000~au.  We then assessed the correspondence with the KDE. The latter is computed on the \ce{HC3N} blue-shifted ($4.65-7.37$~km~s$^{-1}$) velocity map, within a mask shown in black contour in Fig.~\ref{fig:streamline-model-HC3N}. As in the previous section, this mask implies a contamination by the disk and/or the inner envelope at small radii. As we searched for a better constraint of the velocity profile, we varied $v_{r,0}$ across the following values: 0~km~s$^{-1}$, $0.4\,v_{ff,\,0}=0.31$~km~s$^{-1}$ and $v_{ff,\,0}=0.78$~km~s$^{-1}$.

Unlike the primary axis, multiple parameter sets within these ranges match both the projected trajectory on the plane of sky and the KDE around the $(\mathcal{R}_z)$ axis. One of them is shown as an example in Fig.~\ref{fig:streamline-model-HC3N}. However, no satisfying fit was found for the $(\mathcal{R}_o)$ or the $(\mathcal{R}_d)$ axis. The streamline model does not enable us to fully constrain the geometry of the streamer, but it confirms that material is infalling toward the central object in the $(\mathcal{R}_z)$ axis case. Moreover, modeled trajectories fall near the observed SO peaks, supporting the accretion shocks hypothesis discussed in Sect.~\ref{sec:discussion-so}.

\subsection{\texttt{TIPSY} fitting model}

Based on the generalized form of the equations of \citet{mendoza2009}, \texttt{TIPSY}\footnote{\url{https://github.com/AashishGpta/TIPSY}}, standing for Trajectory of Infalling Particles in Streamers around Young stars, is a python package that directly fits the position-position-velocity (PPV) cube to test the infalling nature of observed material \citep{tipsy2024}. \texttt{TIPSY} first generates a curve-like representation of the observed streamer structure and then compares this curve to the theoretically expected trajectories of infalling material. A solution is determined to be the best fit based on its fitting fraction, defined as the fraction of observed values consistent with the theoretical trajectory, and its $\chi^2$ deviation.
The model inputs are:
\begin{itemize}
    \item the PPV cube to fit ;
    \item the mass of the central object $M_\star$, set to 1.7~$M_\odot$ ;
    \item the distance to the object, set to 146~pc ;
    \item the systemic velocity of the source, set to 7.37~km~s$^{-1}$.
\end{itemize}
The computation of the trajectory then enables us to derive various parameters, like the direction of the rotational axis through the angular momentum vector, the kinematic, potential and total energies, or the infalling time. We tested this model to compare its output to the "visual fits" made in Sect.~\ref{sec:streamline-model}.

\subsubsection{Along the primary axis}
\label{sec:tipsy-cs}
We used the full CS PPV cube masked by the 65$\sigma$ level contour from the CS blue-shifted emission (as in Sect.~\ref{sec:streamline-model-CS}), using only emission above a 5$\sigma$ threshold, but the model did not converge (see Fig.~\ref{fig:tipsy-cs-fit-chi2}, fitting fraction of 0.73 with a $\chi^2=47.9$ deviation). As the PV diagrams along the primary axis revealed a more pronounced arm-like structure in the \ce{C2H} emission (see Fig.~\ref{fig:quadrupolar-pv}), we used the \ce{C2H} PPV cube as input, applying the same masking as for CS. However, once again, the model failed to converge, leaving the streamer hypothesis unconfirmed. 

\subsubsection{Along the secondary axis}
\label{sec:tipsy-hc3n}
\begin{figure}
    \centering
    \includegraphics[width=\linewidth]{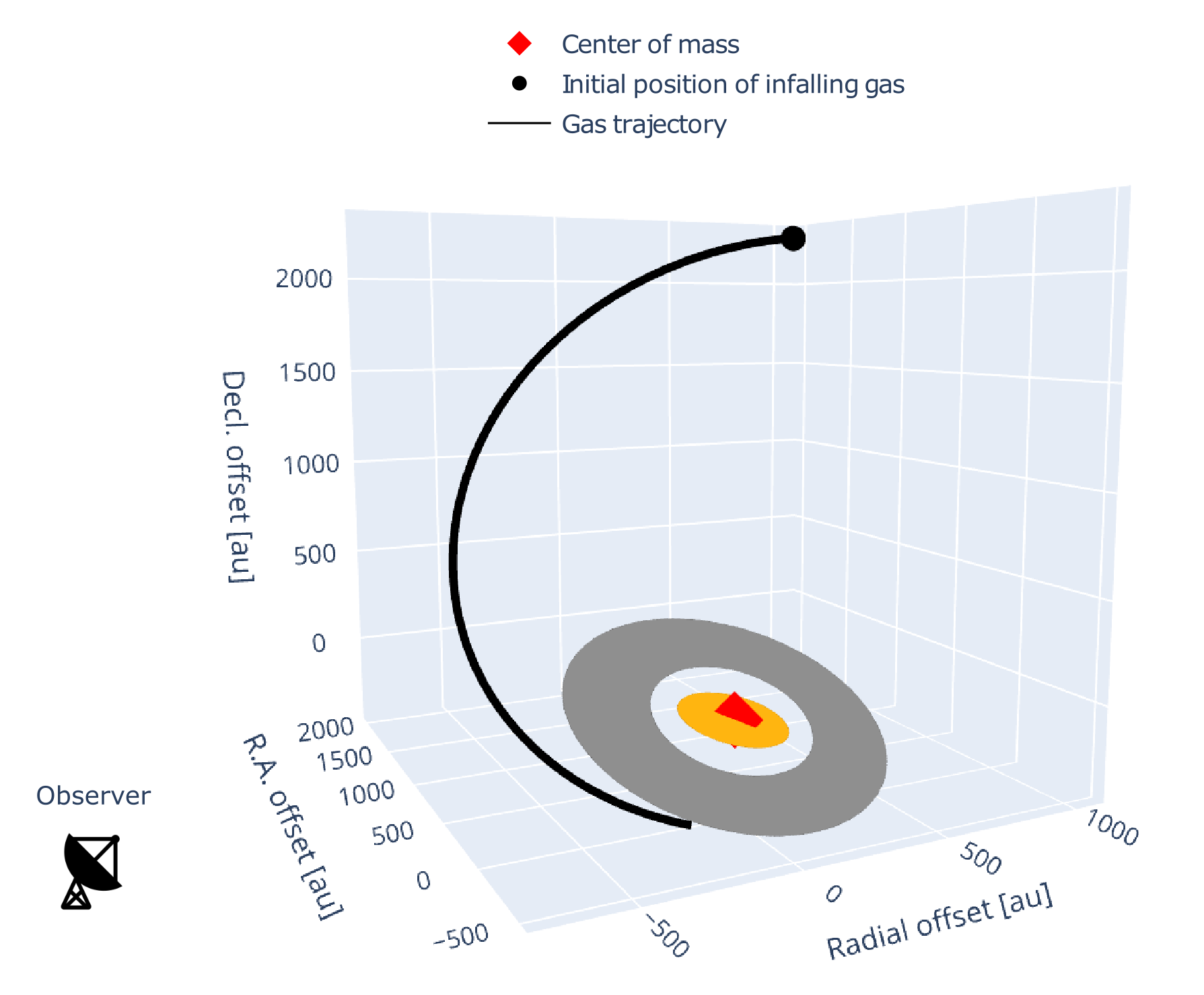}  
    \caption{Trajectory computed by \texttt{TIPSY} from the \ce{HC3N} PPV cube. The outer and intermediate disks are displayed in grey and orange. Their parameters come from \citet{sai2020}.}
    \label{fig:tipsy-hc3n-traj}
\end{figure}

We used the full \ce{HC3N} PPV cube applying the same mask as the one described in Sect.~\ref{sec:streamline-model-HC3N} using only emission above a 5$\sigma$ threshold. The model converged and returns the best model shown in Fig.~\ref{fig:tipsy-hc3n-traj} with the highest fitting fraction (0.97) and the lowest deviation ($\chi^2=5.47$, see Fig.~\ref{fig:tipsy-hc3n-fit-chi2}). The modeled material is coming toward us which is consistent with blue-shifted emission.
Using the same definitions as in Sect.~\ref{sec:streamline-model}, relative to the $(\mathcal{R}_z)$ axis, it gives $r_0=3\,230\pm165$~au, $\theta_0=44.9\pm2.9\degr$~and $\varphi_0=-207.7\pm6.2\degr$. The output rotational axis is $i=-2.7\degr$~and $\text{PA}=-47.9\degr$.

\subsection{Summary and models comparison}
\label{sec:summary-infall-models}

The PV diagrams of \ce{HC3N} and \ce{C2H} along both axes showed a similar arm-like structure with a velocity increase near the disk (see Fig.~\ref{fig:quadrupolar-pv}), reminiscent of infalling material. However, both models along the primary axis are not able to reproduce the elongated blue-shifted emission, challenging this infalling hypothesis discussed in Sect.~\ref{sec:discussion-primary-axis}. It highlights the need of modeling the emission before drawing conclusions based on hypotheses derived from the PV diagrams. For the secondary axis, both models agree on infalling material toward the disk, but multiple parameter sets lead to similar trajectories and velocity profiles projected on the sky's plane that reproduce the observations. 
The main challenge is determining the best parameter sets, especially with the streamline model's requirement for initial conditions. This necessitates constructing a grid in $r_0$, $\theta_0$, $\varphi_0$, and $v_{r,0}$, leading to intensive computations and lengthy analyses for model-observation comparisons.
\texttt{TIPSY} makes things easier by directly estimating initial conditions, computing the accuracy of the estimated set of parameters, displaying fit quality maps (see Fig.~\ref{fig:tipsy-cs-fit-chi2} and \ref{fig:tipsy-hc3n-fit-chi2}), and giving the best trajectory. \texttt{TIPSY} can also provide interesting additional parameters such as the infalling time, the total energy, and the angular momentum for instance \citep{tipsy2024}.

\section{Discussion}
\label{sec:discussion}
\subsection{Late infall}
\subsubsection{Warped disk origin}
\label{sec:warp-origin}
\citet{brinch2007b} suggested that a misalignment between the rotational axes of the envelope and the disk is causing the external warped disk in this source.  
To test this hypothesis, \citet{sai2020} performed an hydrodynamical simulation involving infalling gas, during which the angular momentum of the gas changes direction after a certain time (equivalent to 10 rotation periods of the gas at 200~au). As a result, they successfully formed a warped disk comparable to the one observed in \object{L1489 IRS}. However, \citet{sai2022} measured the specific angular momentum from the envelope to the disk, using \ce{C^18O}~($2-1$), and found their observations to be consistent with models of inside-out collapsing cores conserving their angular momenta, and thus their overall rotational axis. 

The use of the streamline model and \texttt{TIPSY} along the secondary axis enabled us to confirm infalling material from the northeast toward the disk, which is consistent with \cite{sai2022}. The streamer is nested within a larger envelope and is likely connected to the disk. Given that \object{L1489 IRS} is considered to be a late Class I protostar, we see a late infall, like in the Class I object \object{Per-emb-50} \citep{valdivia-mena2022}.
In some hydrodynamical simulations, it has been demonstrated that the interaction of infalling material on an existing disk can result in the formation of a second-generation disk around the initial one. This newly formed disk may exhibit misalignment relative to the inclination of the infalling material's trajectory, all without requiring any prior misalignment between the envelope and the initial disk axis.
In this scenario, the resulting misaligned disk system would be able to survive for more than 100~kyr without aligning with each other \citep{kuffmeier2021}. This phenomenon could explain the warp between the outer ($300-600$~au) and the intermediate disk ($15-200$~au) as illustrated in Fig.~\ref{fig:schematic} and \ref{fig:tipsy-hc3n-traj}. 

An alternative scenario could be a break in the disk due to a binary protostar. However, this would occur between 5$a_B$ and 15$a_B$ (according to the disk viscosity) where $a_B$ is the semi-major axis between the protostars \citep[][Cuello et al. in prep.]{nixon2013,facchini2018,rabago2023arxiv,young2023}. The best resolved images of \object{L1489 IRS} up to date are 1.3mm continuum images with a resolution of 11~au but no binary was resolved \citep{yamato2023}, resulting in two different cases. Either $a_B>11$~au, and the secondary was thus not detected. This could be explained by a very unequal mass binary, or a secondary with no mm emission, but an indirect effect of its presence across the continuum emission would be expected, like the gap seen $\sim$\,30~au \citep{yamato2023}. This gap may indeed have been dug by a planet whose mass would be less than 2.4~$M_\text{Jup}$ \citep{yamato2023}. However, the mass ratio between the two companions would be substantially lower than the minimal 0.2 value needed to break the disk (Cuello et al. in prep.). Or $a_B<11$~au, and the binary is simply not resolved. An upper limit on the break radius $R_\text{break}$ due to the binary is therefore $R_\text{break}\leq165$~au (i.e., $15a_B$) which is less than the gap location modeled by \citet{sai2020}, but could explain the gap identified around 30~au \citep{yamato2023}. Assuming $R_\text{break}=30$~au, it gives $a_B=3-6$~au which is close to the estimation of \citet{covey2006} of $a_B=2.4$~au based on theoretical orbital motions for $M_\star=1.6~M_\odot$. 

Stellar encounters could also produce similar effects \citep{cuello-review2023}. The presence of an external companion to the system, following a periodic orbit misaligned with the intermediate disk, could also results to such geometries in Class II disks \citep[whose accretion rates, and viscosities, are different from Class 0/I stages, and could impact the lifetime of structures,][]{long2021, gonzalez2020,nealon2020b}. However, to the best of our knowledge, no companion has been detected in the L1489 IRS system so far. Hence, the late infall theory is the most compelling explanation for the warped of the outer disk.

Finally, it's worth noting an intriguing characteristic angle of this system. Indeed, 14\degr~separates the mean P.A. of the observed streamer (P.A.$=40\degr$) from the outer disk \citep[P.A.$=54\degr$,][]{sai2020}, 13\degr~the outer disk from the intermediate disk \citep[P.A.$=67\degr$,][]{yamato2023}, and 15\degr~the intermediate disk from the inner disk \citep[P.A.$=82\degr$,][]{yamato2023}. This $\sim15\degr$ angle may reflect a preferred dynamical configuration adopted by the system to stabilize itself, that could originate from an angular momentum transfer between the different disks, with the outflow constraining the P.A. variation of the most inner disk.

\subsubsection{Core-disk connection} 
\begin{figure}
    \centering
    \includegraphics[width=\linewidth]{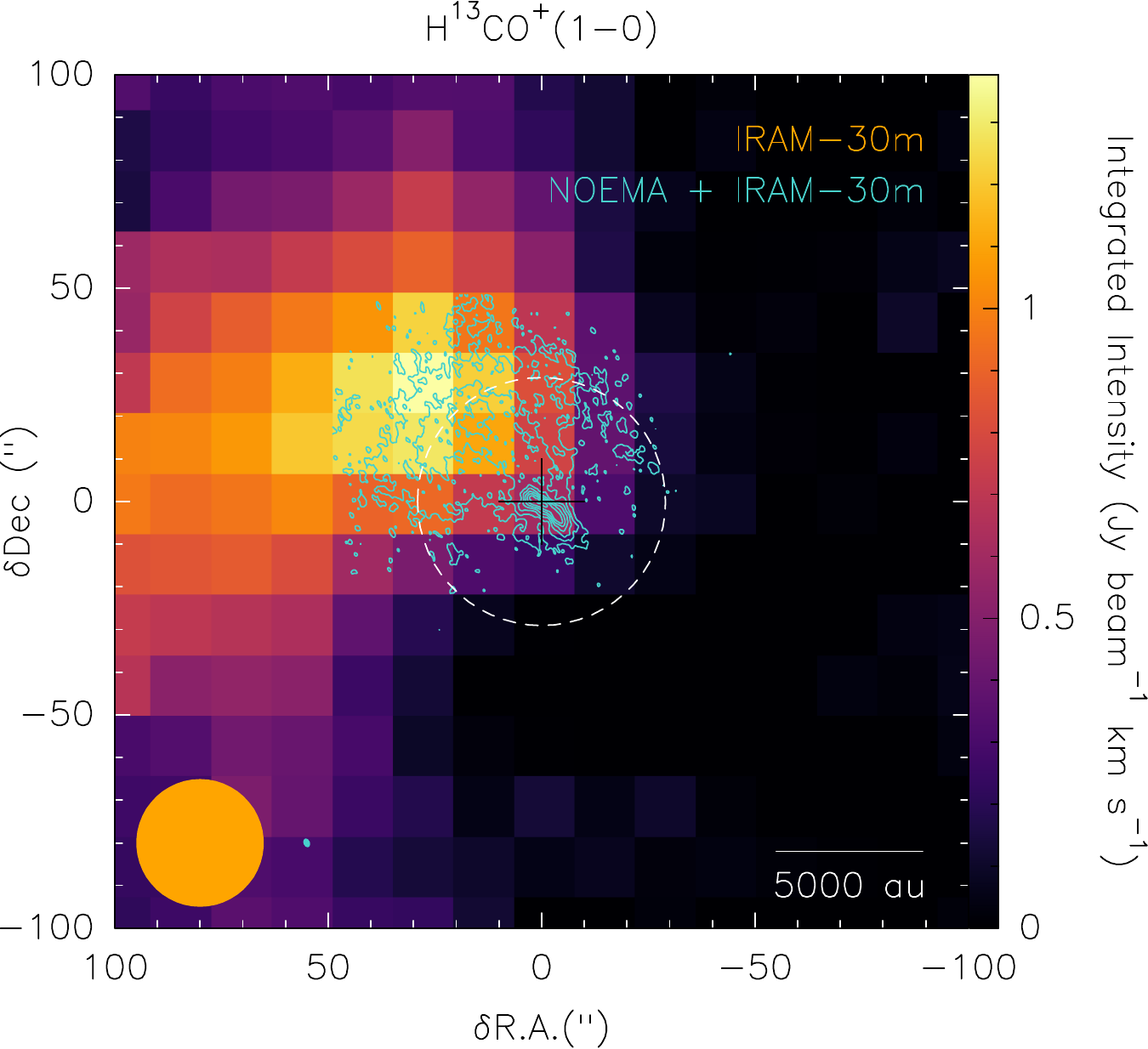}
    \caption{Integrated intensity maps of the \ce{H^13CO+} emission from the IRAM-30m (background) and combined (IRAM-30m + NOEMA) data (contours). The contour levels are 5$\sigma$ to 55$\sigma$ by 5$\sigma$ steps where $\sigma=1.75$ mJy~beam$^{-1}$~km~s$^{-1}$. The dashed white line shows NOEMA's primary beam. The beams are displayed in the lower left corner. The scale bar on the bottom right corner indicates 5\,000~au.}
    \label{fig:h13cop-30m-merged}
\end{figure}

The PV diagram along the streamer shows a velocity gradient from the nearby prestellar core $v_\text{LSR}=6.6-6.8$ km~s$^{-1}$ \citep{wu2019}, to the disk $v_\text{LSR}=7.37$ km~s$^{-1}$ \citep[][see. Fig.~\ref{fig:quadrupolar-pv}]{yamato2023}. Moreover, the \ce{H^13CO+} emission on the northeast, as shown in Fig.~\ref{fig:bit-h13cop} from combined data, is localized at the edge of the core where \ce{H^13CO+} is detected in the IRAM-30m data (see Fig.~\ref{fig:h13cop-30m-merged}). The latter shows an emission peak outside the primary beam of NOEMA, suggesting that the emission is extended further than what the combined data show. Similar conclusions are drawn from the other tracers of large structures in this survey, such as \ce{HCO^+}, \ce{HCN}, \ce{C2H} and \ce{c-C3H2}.
Therefore, the link in gas between the prestellar core and the YSO is very likely to be present, which is consistent with the modeling of \citet{brinch2007a}. This would imply that the prestellar core is feeding the protostar, offering interesting prospects to investigate the chemistry from the core to the disk. 
A similar conclusion was reached for the Class I protostar \object{Per-emb-53} localized in the dense core region of \object{Barnard 5} \citep{valdivia-mena2023}.

L1489 IRS could have formed within the core and then migrated outside of it, as suggested by its position at the edge of the core \citep{brinch2007a}. \citet{adams2001} estimated a gravitational tide radius of the Pleiades cluster of $\sim\,6\degr$ on the sky's plane. This system is found to be within this radius, and could have belong to the Pleiades rather than Taurus \citep{rebull2011}. The YSO is localized on the edge of its parental cloud and toward the Pleiades cluster which might attracted L1489 IRS, giving a potential explanation to its migration from the core.
To confirm the link between the prestellar core and the protostellar object, mosaic observations of this connection are required, as the extent of the emission goes beyond the primary beam of NOEMA. 

\subsection{SO, a shock tracer}
\label{sec:discussion-so}
As pointed out in Sect.~\ref{sec:secondary-axis} and Fig.~\ref{fig:streamer-shocks}, we see a clear spatial correlation between the SO emission peaks and the base of the emission of infall tracers. A similar observation has been made between \ce{C^18O}~$(2-1)$ and higher frequency lines of SO \citep{yen2014,yamato2023}. Accretion shocks with slow velocities (around 1 km~s$^{-1}$) likely due to streamers have been observed in similar Class I/II disks like \object{DG Tau} and \object{HL Tau} along the streamers \citep{garufi2022}. Using a ring model for \object{L1489 IRS}, \citet{yen2014} were able to reproduce the PV diagram of the SO~$(5_6-4_5)$ emission. This is consistent with core collapse simulations using non-ideal MHD effects coupled with chemistry, which predict the formation of a streamer that shocks the disk, if the streamer lands in the equatorial plane of the disk \citep{mauxion2024arXiv}. From the \texttt{TIPSY} best-fit trajectory, considering typical Class I disks gas scale heights \citep{podio2020a} we estimate the incidence angle between the streamline and the external layer of the disk on the side of the streamline to be 8.8\degr for $z/r=0.2$, and 3.1\degr for $z/r=0.3$. Using $z/r\geq0.4$, the streamline falls into the equatorial plane of the disk. Therefore, accretion shocks in the disk due to the streamer is the most likely explanation.

Concerning the offset SO blob described in Sect.~\ref{sec:secondary-axis},
it is further away and not in contact with the disk (i.e., even with the most external disk). A similar configuration is seen in the \object{Oph-IRS 44} Class I system with an offset region of \ce{SO_2} located inside a streamer at $\sim400$~au of the protostar \citep{ArturdelaVillarmois2022}. According to the low temperature of this \ce{SO_2} spectral line ($E_\text{up}=36~K$), \citet{ArturdelaVillarmois2022} classified this region as a \ce{SO_2} knot. The SO blob could also just be a knot within the red-shifted infall along the secondary axis in our case. However, the upper energies of our SO lines are lower than 36~K and yet the most natural explanation of the other two emission peaks is an accretion shock as explained above. Hence, the same phenomenon could happen here, with an accretion shock occurring onto the dense envelope close to the disk, rather than directly onto the disk itself. 

We also note that the integrated intensity map of the SO~$(2_3-1_2)$, shown in Fig.~\ref{fig:mom0-maps}, displays an elongated structure on the north that lies on the edge of the northeastern envelope, traced by \ce{H^13CO^+} for instance. In \object{HH 212}, a $\sim$\,300~au rotating flow probed in SO was identified and suspected to be associated to an MHD disk wind \citep{tabone2017,lee2021}. However, in our case, this emission is more than ten times longer and is not symmetric with respect to the \ce{HCO+} outflow axis, making this interpretation unlikely. Instead, it could be attributed to a recollimation shock, assuming a launching radius of the molecular outflow $<5$~au \citep{jannaud2023}, which is consistent with resolved Class 0 and I observations \citep{pascucci2023}. Another explanation could be that this elongated SO structure reflects a shearing layer at the interface between the infalling material from the northeast and the outflow, due to opposite directions of their velocities \citep{cunningham2005,tabone2018}.

\subsection{Primary axis' flow nature}
\label{sec:discussion-primary-axis}
We discuss here the nature of the blue-shifted emission along the primary axis, as raised by \citet{ohashi1996} who could not distinguish between infall or outflow due to a too low spatial resolution with the NMA ($\sim$\,8\arcsec).

\ce{HC3N}, \ce{C2H}, and CS PV diagrams along this axis (see Fig.~\ref{fig:quadrupolar-pv}) suggest infalling material. This inference is further supported by the presence of core material along this direction. This aligns with CS and SO observations toward YSOs in the Perseus star-forming region, which have identified these molecules as tracers of infalling material \citep{ArturdelaVillarmois2023}. Nevertheless, our infalling models, as described in Section~\ref{sec:summary-infall-models}, do not align with the observations.

Furthermore, the blue-shifted emission along the primary axis shows a bending pattern toward the blue lobe of the outflow, situated at the edge of the cavity, resembling a configuration observed in the case of the Class I object \object{Oph IRS 63} \citep{Flores2023}. 
This raises the possibility that the outflow could perturb a free-fall trajectory as the one modeled in our study. 
Alternatively, the outflow could also contribute to feed this infalling material.
However, confirming this hypothesis would require a comprehensive chemical analysis, involving the comparison of molecular abundance ratios derived from radiative transfer models, which is beyond the scope of this paper.

Considering, the blue-shifted emission along the primary axis as additional infalling material implies the presence of at least three distinct infalls toward the disk and one outflow for this Class I object. Interestingly, a similar configuration with multiple streamers and an outflow has been recently observed in an other object at an earlier stage, the Class 0 protostar \object{IRAS 16544-1604} \citep{kido2023}. 
Additional investigations, involving a broader sample of Class 0 and Class I systems, are needed to further assess the relationship between multiple infalls and a single outflow.

On the other hand, the outflow hypothesis is also a possibility. Optical images from the HST show a structure reminiscent of an outflow cavity in the southwest direction \citep{padgett1999}. This pattern matches the \ce{H2}~$(1-0)$~S(1) and K-band emission of \citet{lucas2000} along this axis, with V-shape structures indicative of high-velocity material movements, characteristic of outflows. Additionally, the red-shifted \ce{^12CO}~$(2-1)$ emission traces an elongated structure toward the northwest \citep{yen2014}, remarkably coinciding with the P.A. of the blue-shifted emission detected with CS, HCN and SO (see Sect.~\ref{sec:primary-axis}). These components may correspond to the blue and red lobes of the hypothetical second outflow. Moreover, multiple HH objects are detected in S[II] at large distances from the YSO (0.4\arcmin~to 4.6\arcmin), roughly aligned along the same P.A. \citep[see Fig.~\ref{fig:herschel-HH},][]{gomez1997}. The unstructured distribution of these objects on larger scales suggests that they may originate from a YSO's outflow whose direction changes erratically over time \citep{gomez1997}.  ,
This outflow would have been observed with \ce{^12CO}~$(1-0)$ \citep{myers1988} 
but this emission seems to rather come from the large cavity of the parental cloud B207 traced in optical and infrared bands \citep[][see Fig.~\ref{fig:herschel-HH} and]{togi2017}. The positions of the HH objects also seem to follow this large cavity whose origin remains unknown and which could also be the origin of these objects. If this elongated blue-shifted emission is indeed an outflow, the YSO has two outflows, confirming the binarity of the system. These outflows would also define a cavity where material could fall onto the disk, which is consistent with the \ce{C2H} and \ce{c-C3H2} emissions (see Fig.~\ref{fig:mom0-maps}). In this scenario, it is important to note that this second outflow would not be accurately aligned with any of the system's disks.

\subsection{Bubbles origin} 
\label{sec:bubbles-discussion}
Two bubbles around the YSO are seen with the dense molecular tracers \ce{HCO+}, CS and HCN (see Fig.~\ref{fig:bubbles}). The superposition of these three different tracers shows the exact same structure, suggesting that these structures are likely more related to physical conditions than chemical processes. As discussed in Sect.~\ref{sec:warp-origin}, the extent of interaction due the potential binarity of the system is limited to $\sim$\,100~au, clearly less than the $\sim$\,10\,000~au width of the bubbles. Similar arc-like or ring structures have been found around YSOs with a range between $\sim$\,100~au and $\sim$\,10\,000~au width, but are either attributed to streamers \citep[e.g.,][]{tokuda2014,yen2019,alves2020,pineda2020,garufi2022,murillo2022,valdivia-mena2022,mercimek2023}, outflows seen face-on \citep{fernandez-lopez2020,harada2023} or shocked gas from an interaction with an outflow \citep{sai2023}. However, a $\sim$\,7\,000~au width bubble spaced $\sim$\,5\,000~au from the Class I protostar \object{CrA-IRS 2} has recently been reported, and is likely a magnetic bubble created following a removal of magnetic flux \citep{tokuda2023}.
The bubbles in Fig.~\ref{fig:bubbles} are indeed reminiscent of star-forming simulation outputs from core collapse models, where expanding magnetic bubbles, that are regions where the magnetic pressure dominates the thermal pressure, dig cavities around the YSO, either for low-mass \citep{hennebelle2020, tu2023arxiv} or high mass stars \citep{mignon-risse2020, mignon-risse2021a, mignon-risse2021b}. The material is thus not able to enter the magnetic bubble and thus accumulates at its edges. In the low-mass regime, with the collapse of a one solar mass core, the width of these bubbles is $\sim$\,250~au \citep{hennebelle2020}. In the high-mass regime, with the collapse of a hundred solar masses core, the width of these bubbles is $\sim$\,2\,000~au \citep{mignon-risse2021b}. It thus appears the higher the mass is, the wider the bubbles are, but the strength of the magnetic field could counteract this effect: the higher the magnetic field is, the wider the bubbles are \citep{hennebelle2020}. A very preliminary study with the \texttt{RAMSES} code \citep{teyssier2002} showed that we could retrieve the same $\sim$\,10\,000~au width we observe with the $1.7\,M_\sun$ of the system. However, further in-depth studies would be necessary to confirm this preliminary findings, offering interesting prospects. We thus hypothesize the bubbles to be the witness of the magnetic flux removal of the star(s) in this object. To confirm their presence, since they extend beyond NOEMA's primary beam, mosaic observations are required.

\section{Conclusions}
\label{sec:conclusions}

We present a new NOEMA 3mm molecular mapping survey of \ce{HCO^+}, \ce{H^13CO^+}, \ce{HCN}, \ce{H^13CN}, \ce{CS}, \ce{SO}, \ce{C2H}, \ce{c-C3H2} and \ce{HC3N} toward the Class I protostar \object{L1489 IRS}. Our main conclusions are summarized as follows:

\begin{enumerate}
  \item We identified a large streamer ($\sim3000$~au) that stands out in \ce{HC3N}, \ce{C2H} and \ce{c-C3H2} 3mm emission. It is compatible with a generic streamline model but multiple set of parameters are consistent with the emission, preventing us from finely constraining its geometry. This streamer likely lands onto the disk, resulting in accretion shocks seen in the \ce{SO} emission ;
  
  \item Most of the molecular tracers exhibits extension toward the northeast, coinciding with the location of the prestellar core. The observed streamer is likely coming from this core, \object{L1489}, suggesting a gas bridge supplying material from the core to the YSO ;
  
  \item The external warped disk is likely due to the late infall around the YSO, as a binary would not warp the disk that far in radius ;
  
  \item The nature of emission along the primary axis is not clear. It cannot be well fitted by streamers models. That could be due to the complexity of the environment, where an interaction between the infall and the bipolar outflow may have occurred. The PV diagrams suggest infalling material while the HH objects and the geometry of the system suggest an outflow. In the latter case, that would be a direct evidence of a binary system with two outflows ;

  \item We identified two bubbles with a $\sim$\,10\,000~au width in the dense gas tracers \ce{HCO^+}, CS and HCN. An interesting hypothesis would be that their origin is linked to the magnetic flux removal of the protostar(s), as observed in simulations and recent observations. 
\end{enumerate}

To confirm the origin of the large streamer in the prestellar core, and thus its bridge-like connection with the protostellar object, as well as the presence of the bubbles, mosaic observations are required, as the extent of the emission goes beyond the primary beam of NOEMA. If the gas bridge is indeed present, it would enable interesting prospects to investigate the chemical relationship from the core to the YSO through the streamer. It is worth noting that this YSO is a source of interest to dig the question of the interstellar heritage, as it is localized in its parental molecular cloud, offering a view of the time evolution of the chemistry through the look of the different spatial scales. Regarding the bubbles, further in depth studies with numerical simulations are needed to better constrain their origin. Finally, the present study increased the likelihood of a potential second outflow in the L1489 IRS system highlighting the possibility of a binary system, which would also require further observations to be confirmed.

\begin{acknowledgements}
  The authors thank the anonymous referee for the interest and valuable comments that helped to improve paper.
  The authors also thank the IRAM staff for their invaluable work making these observations possible. M.T. and R.L.G. would like to thank Arancha Castro-Carrizo for her advices with the data calibration, Jérôme Pety for his help with the data reduction, and Patrick Hennebelle, Geoffroy Lesur and Laure Bouscasse for useful discussions.
  This work was supported by the Programme National “Physique et Chimie du Milieu Interstellaire” (PCMI) of CNRS/INSU with INC/INP co-funded by CEA and CNES. 
  This work is based on observations carried out under project numbers 184-20, S20AH and W20AJ (PI: R. Le Gal), with the IRAM-30m and IRAM Interferometer NOEMA. IRAM is supported by INSU/CNRS (France), MPG (Germany) and IGN (Spain). This work has benefited from the Core2disk-III residential program of Institut Pascal at Université Paris-Saclay, with the support of the program “Investissements d’avenir” ANR-11-IDEX-0003-01.
  Support for C.J.L. was provided by NASA through the NASA Hubble Fellowship grant No. HST-HF2-51535.001-A awarded by the Space Telescope Science Institute, which is operated by the Association of Universities for Research in Astronomy, Inc., for NASA. 
  This project has received funding from the European Research Council (ERC) under the European Union Horizon Europe programme (grant agreement No. 101042275, project Stellar-MADE). 
  This research made use of \texttt{astropy} \citep{astropy2013,astropy2018,astropy2022}, \texttt{GILDAS} \citep{gildas}, \texttt{gofish} \citep{gofish}, \texttt{matplotlib} \citep{matplotlib2007}, \texttt{numpy} \citep{numpy2020}, \texttt{proplot} \citep{proplot}, \texttt{pvextractor} \citep{pvextractor}, \texttt{scipy} \citep{scipy2020} and \texttt{TIPSY} \citep{tipsy2024}.
\end{acknowledgements}

\bibliographystyle{aa}
\bibliography{refs}

\begin{appendix}
\section{Channel maps}
\label{appendix:channel-maps}

\begin{figure*}
    \centering
    \includegraphics[width=\textwidth]{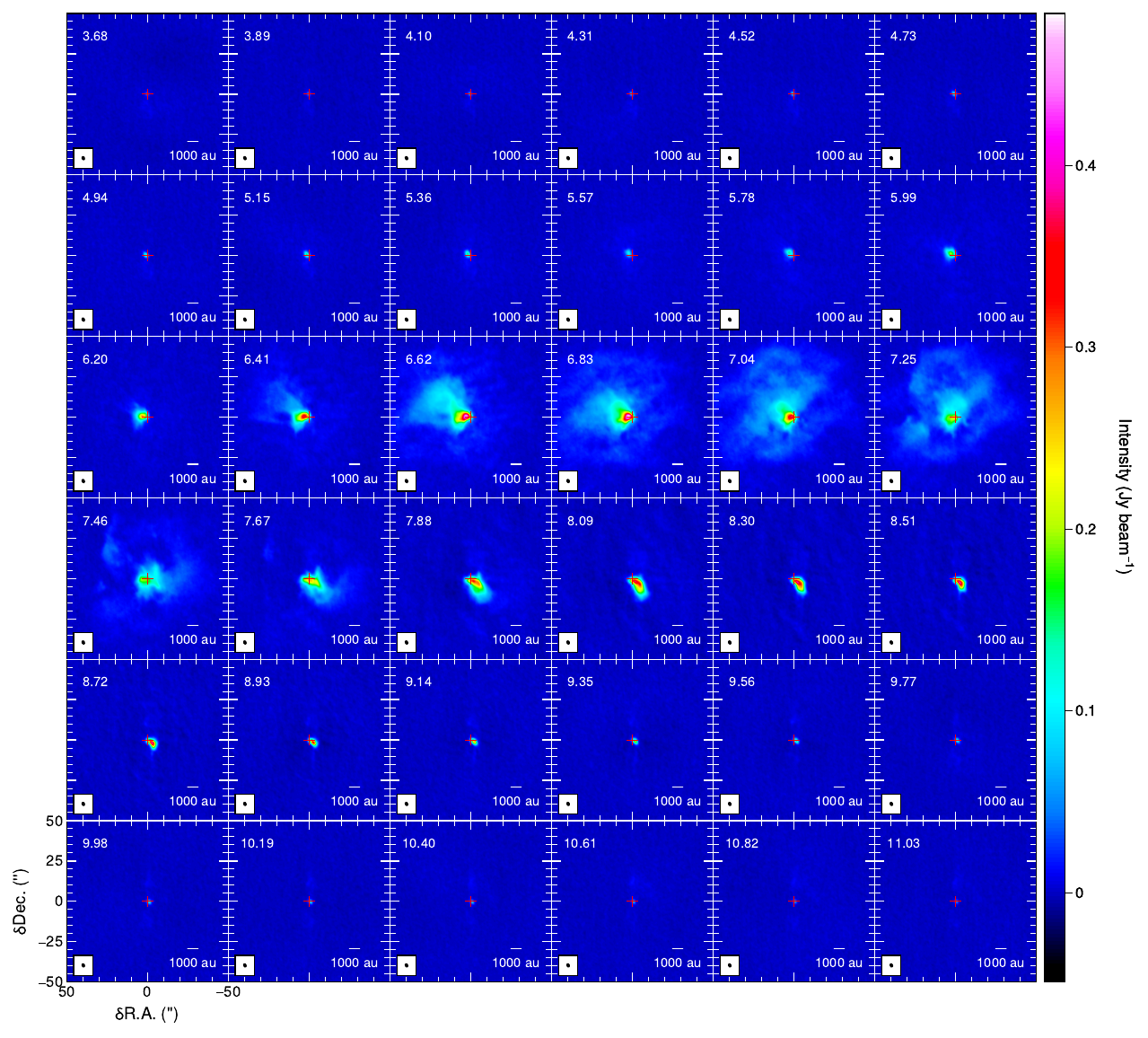}  
    \caption{Velocity channel maps of the \ce{HCO^+} $J=1-0$ emission before primary beam correction. The channel velocity in km$\cdot$s$^{-1}$ is shown in the upper left corner of each panel, while the synthesized beam is shown in the lower left corner. Red crosses show the (0\arcsec, 0\arcsec) position.}
    \label{fig:bit-hcop}
\end{figure*}

\begin{figure*}
    \centering
    \includegraphics[width=\textwidth]{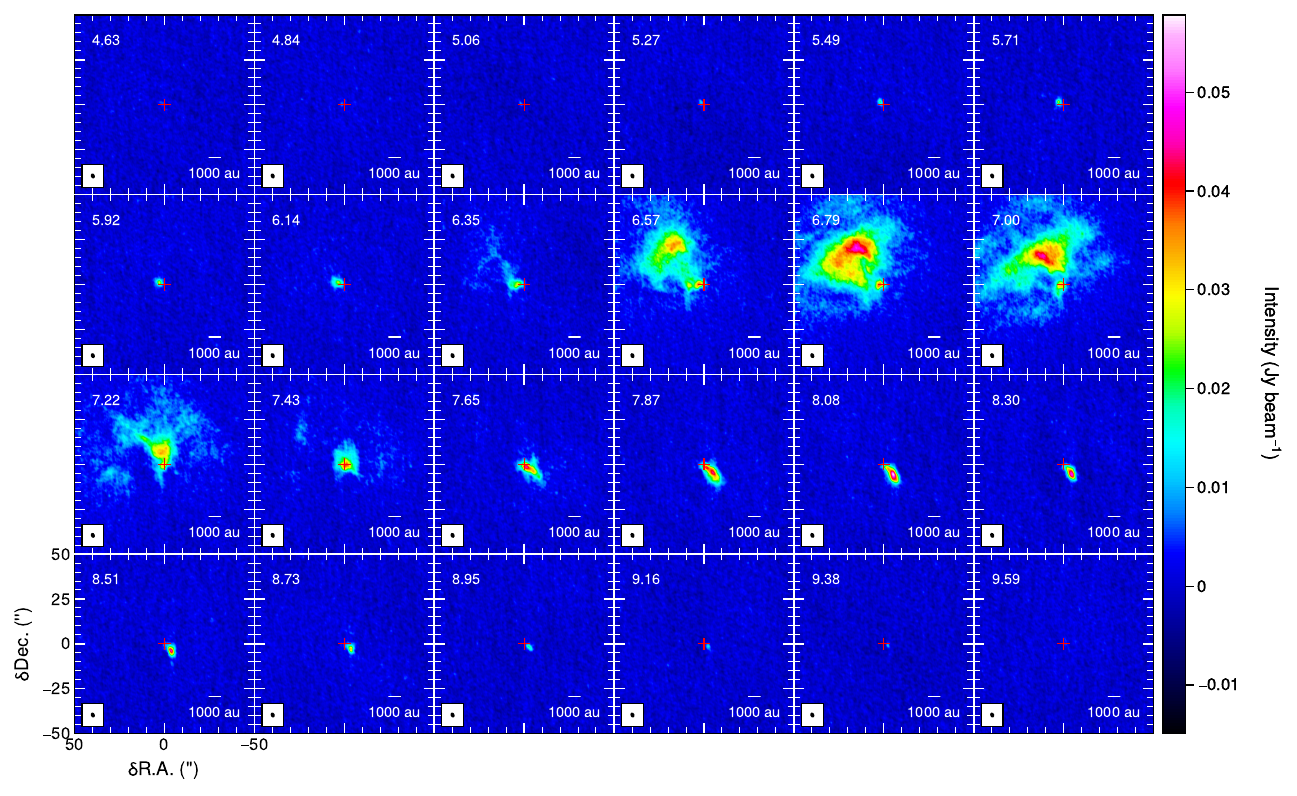}  
    \caption{Same as Fig.~\ref{fig:bit-hcop} but for the \ce{H^13CO^+} $J=1-0$ emission.}
    \label{fig:bit-h13cop}
\end{figure*}

\begin{figure*}
    \centering
    \includegraphics[width=\textwidth]{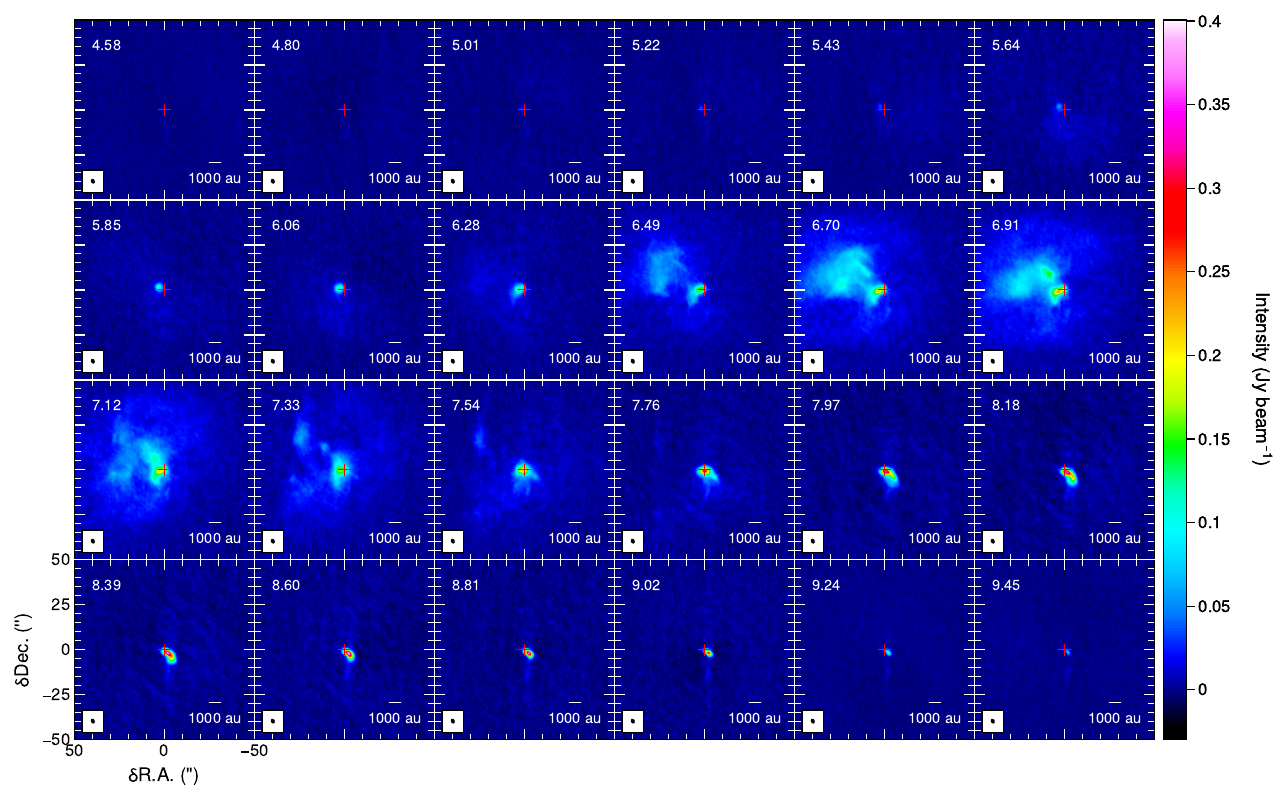}  
    \caption{Same as Fig.~\ref{fig:bit-hcop} but for the \ce{HCN} $J=1-0$ emission. The three hyperfine components are stacked.}
    \label{fig:bit-hcn}
\end{figure*}

\begin{figure*}
    \centering
    \includegraphics[width=\textwidth]{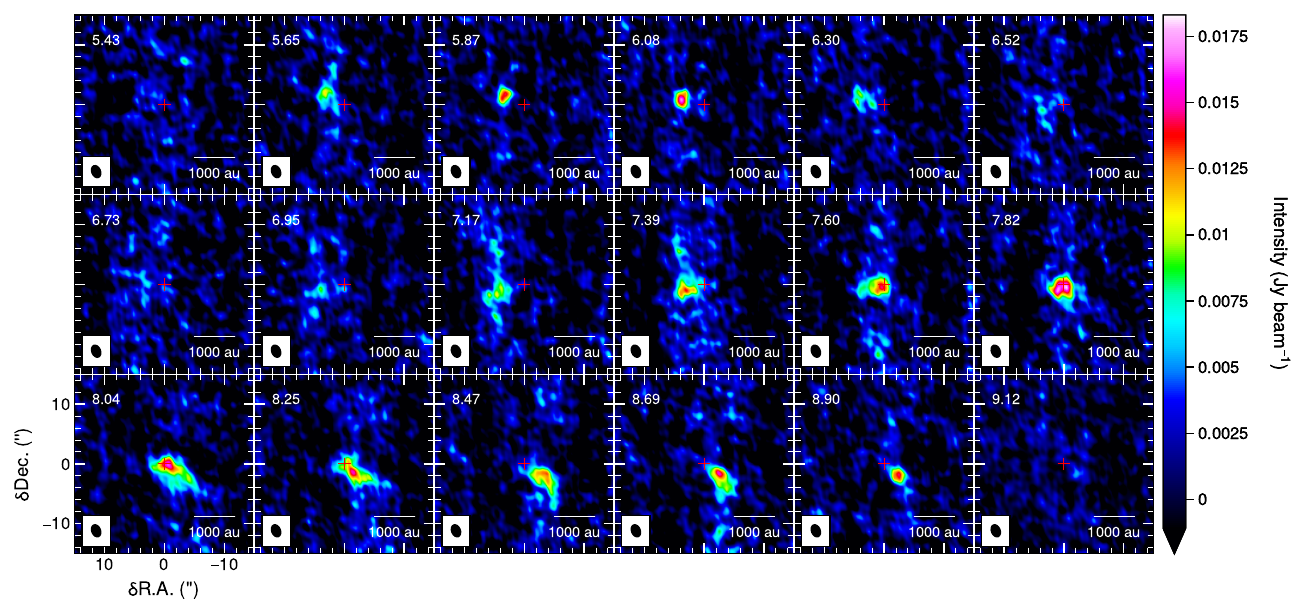}  
    \caption{Same as Fig.~\ref{fig:bit-hcop} but for the \ce{H^13CN} $J=1-0$ emission. Only the $J=1_2-0_1$ and the $J=1_1-0_1$ lines are stacked as the $J=1_0-0_1$ line is not detected. Note that the scale is different.}
    \label{fig:bit-h13cn}
\end{figure*}

\begin{figure*}
    \centering
    \includegraphics[width=\textwidth]{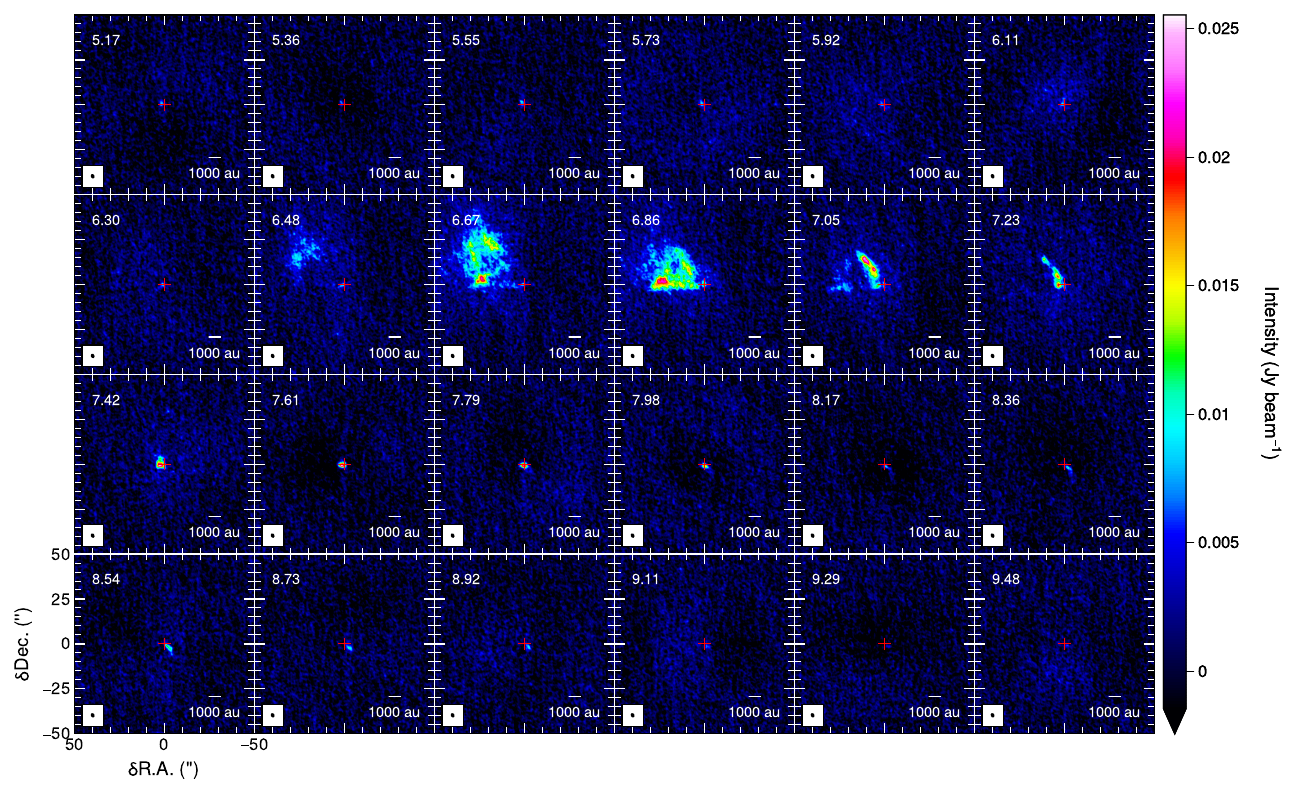}  
    \caption{Same as Fig.~\ref{fig:bit-hcop} but for the \ce{HC3N} $J=11-10$ emission.}
    \label{fig:bit-hc3n}
\end{figure*}

\begin{figure*}
    \centering
    \includegraphics[width=\textwidth]{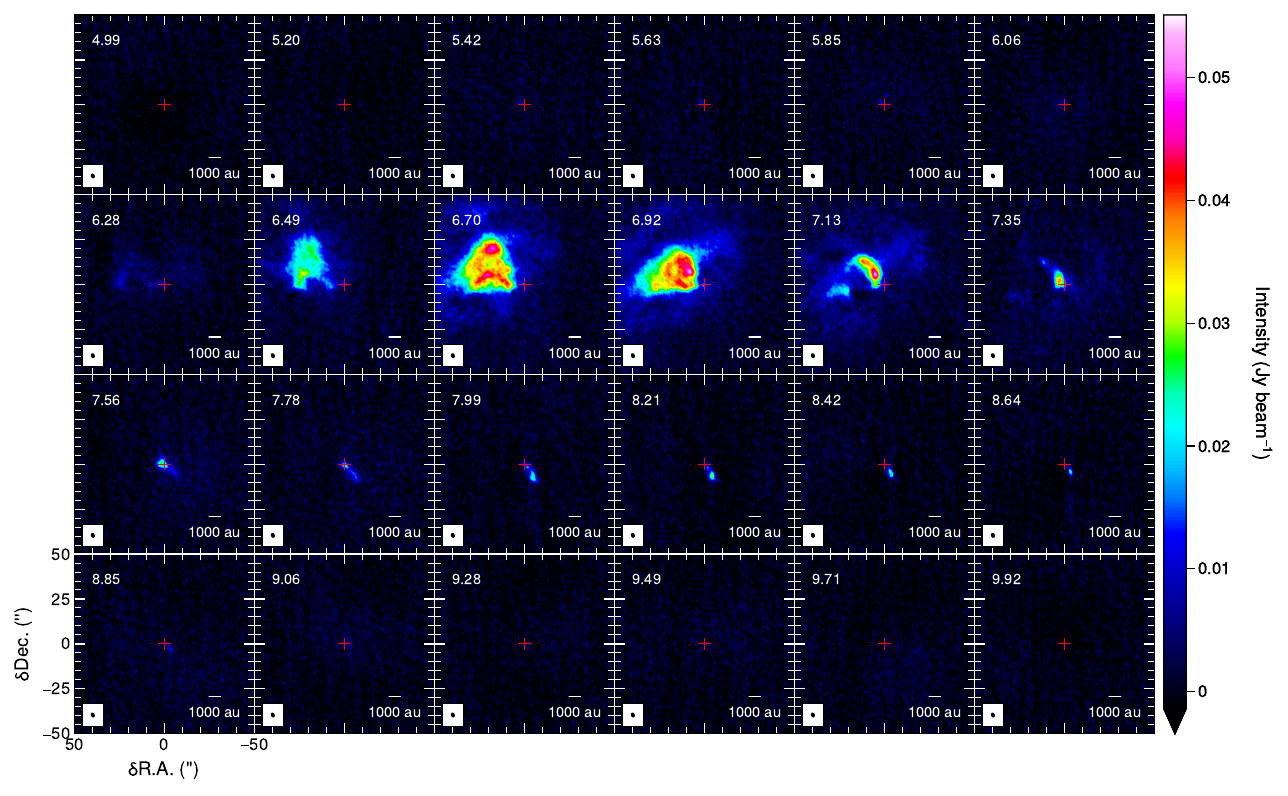}  
    \caption{Same as Fig.~\ref{fig:bit-hcop} but for the \ce{C2H} $J_{N,F}=1_{1.5,2}-0_{0.5,1}$ emission.}
    \label{fig:bit-c2h}
\end{figure*}

\begin{figure*}
    \centering
    \includegraphics[width=\textwidth]{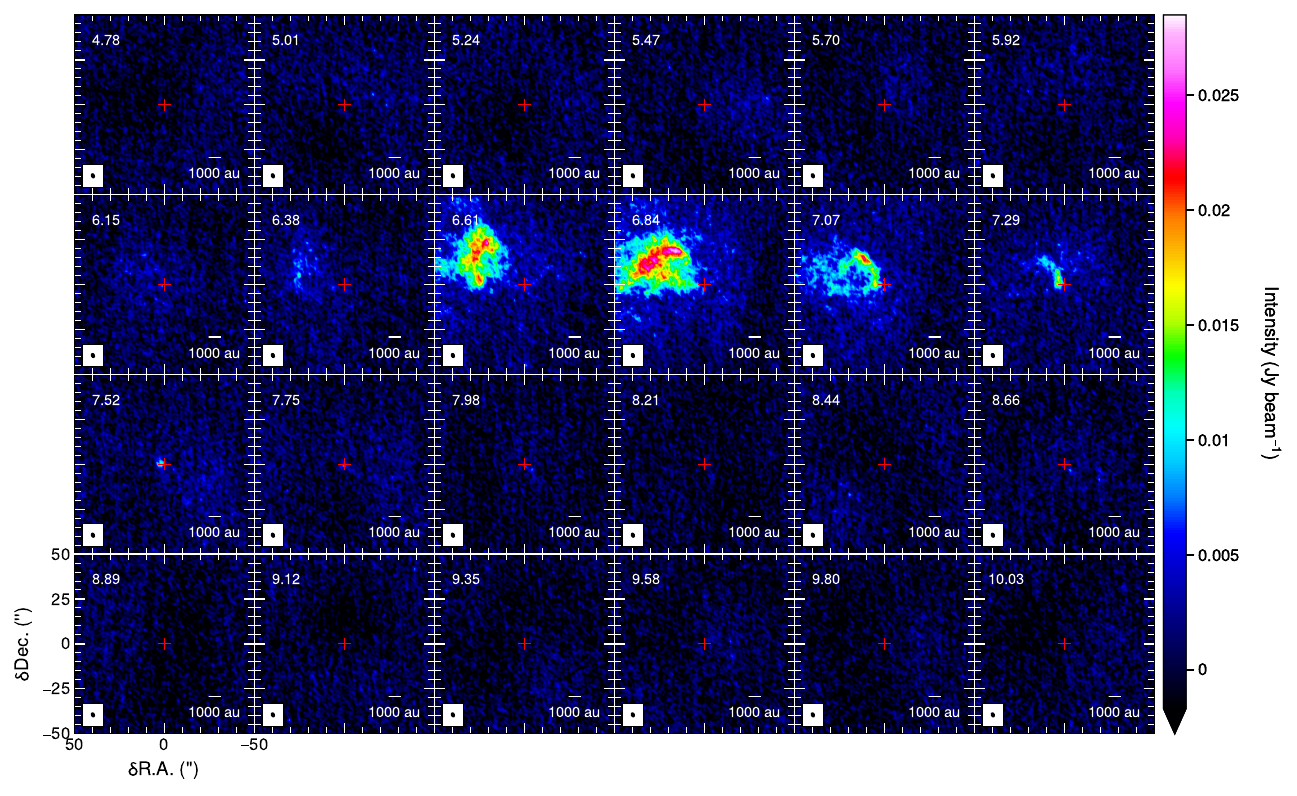}  
    \caption{Same as Fig.~\ref{fig:bit-hcop} but for the \ce{c-C3H2} $J_{Ka,Kc}=2_{0,2}-1_{1,1}$ emission.}
    \label{fig:bit-cc3h2}
\end{figure*}

\begin{figure*}
    \centering
    \includegraphics[width=\textwidth]{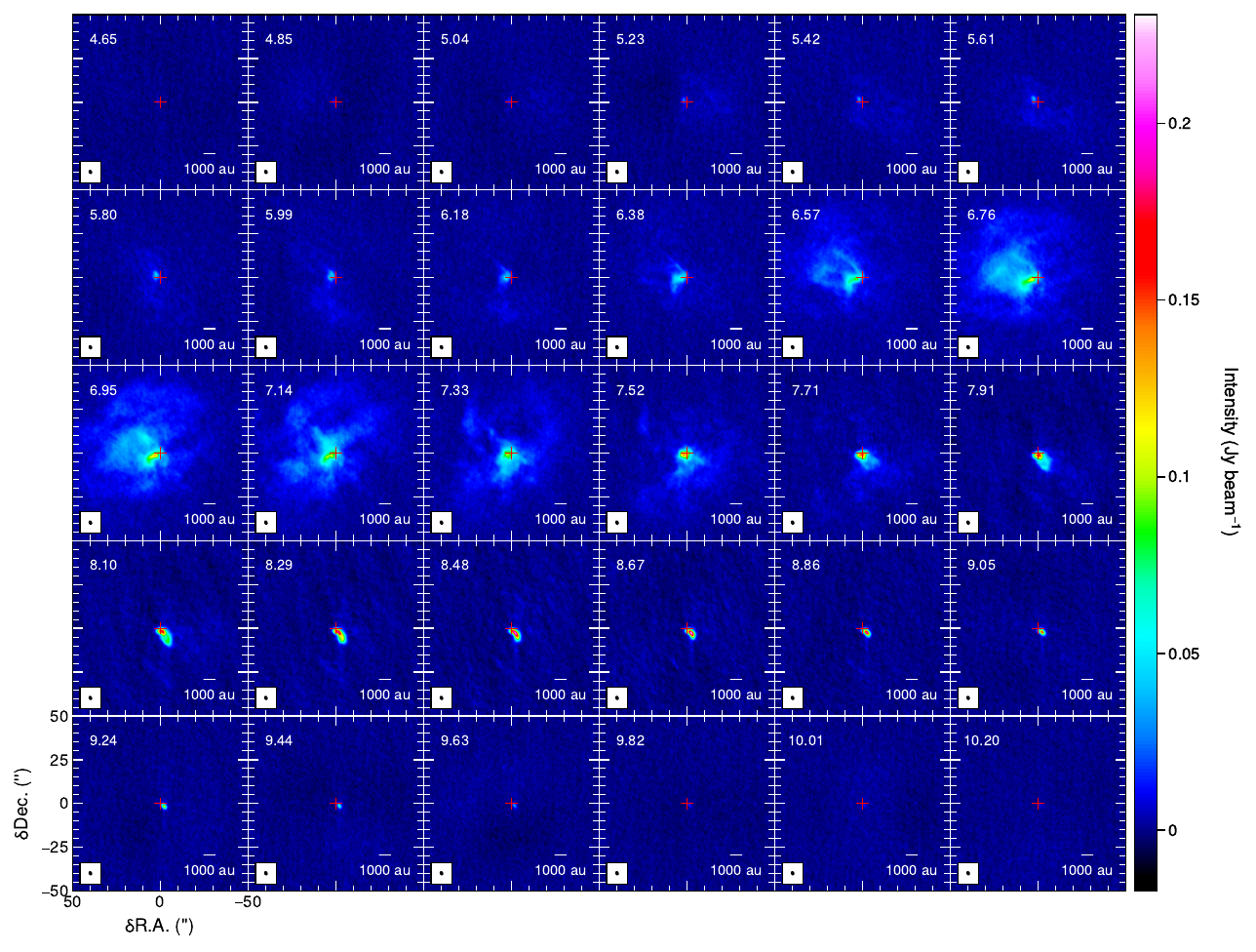}  
    \caption{Same as Fig.~\ref{fig:bit-hcop} but for the \ce{CS} $J=2-1$ emission.}
    \label{fig:bit-cs}
\end{figure*}

\begin{figure*}
    \centering
    \includegraphics[width=\textwidth]{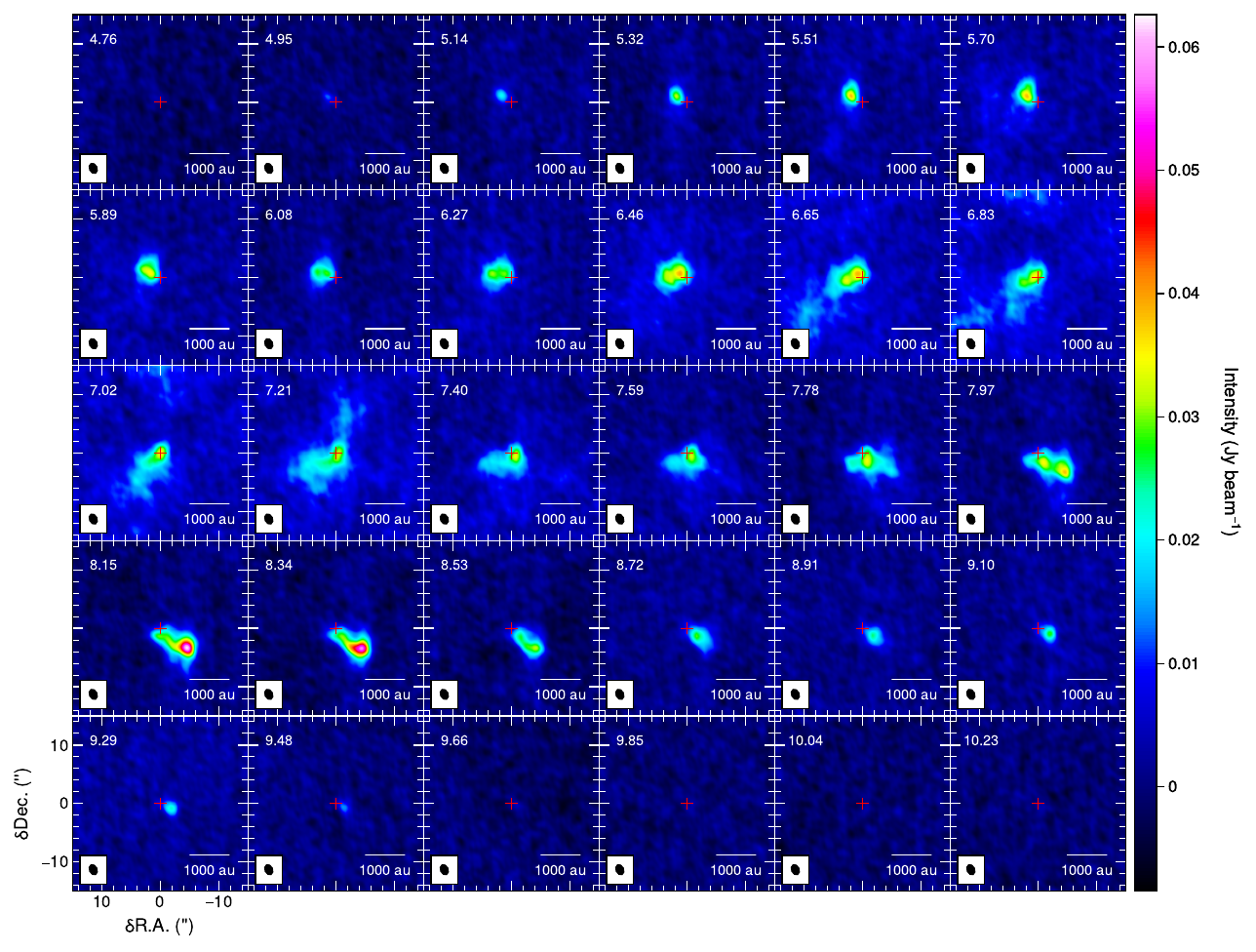}  
    \caption{Same as Fig.~\ref{fig:bit-hcop} but for the \ce{SO} $J_N=2_3-1_2$ emission. Note that the scale is different.}
    \label{fig:bit-so2312}
\end{figure*}

\begin{figure*} 
    \centering
    \includegraphics[width=\textwidth]{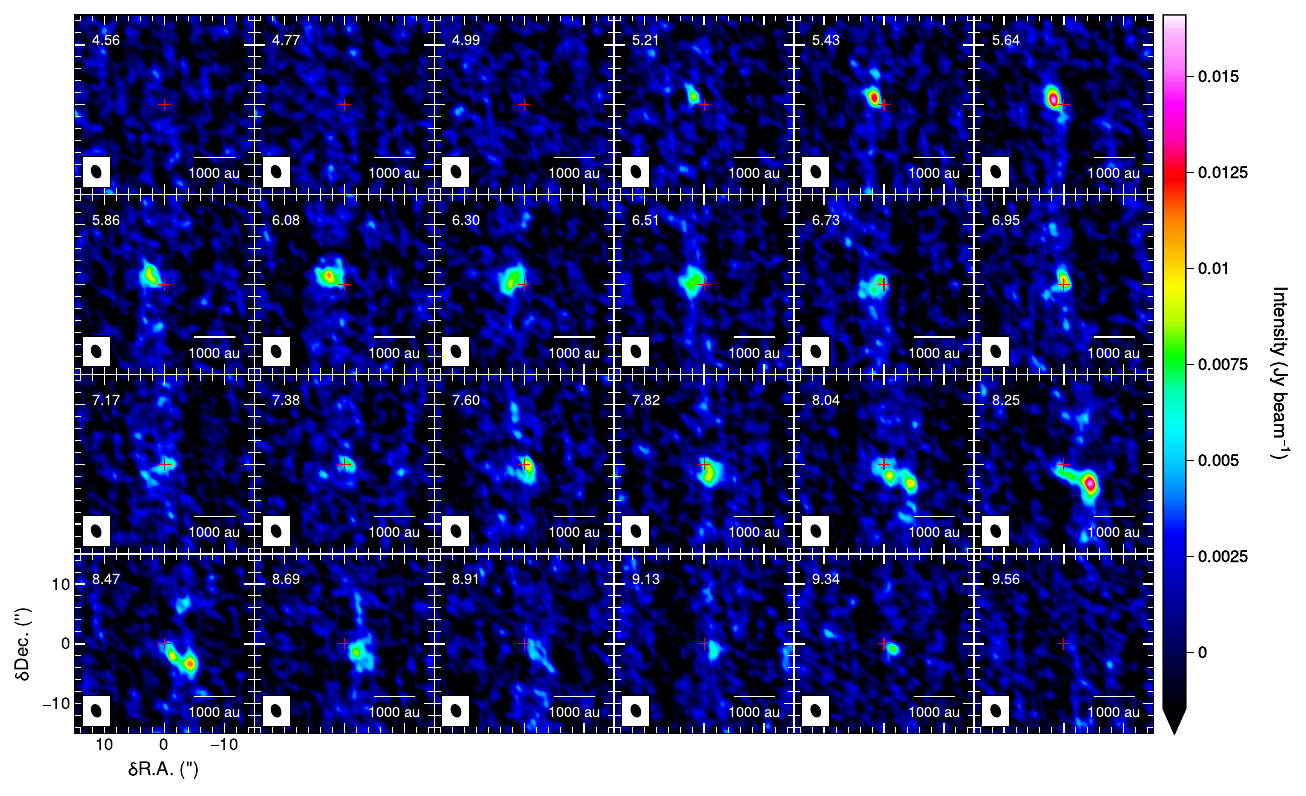}  
    \caption{Same as Fig.~\ref{fig:bit-so2312} but for the \ce{SO} $J_N=2_2-1_1$ emission.}
    \label{fig:bit-so2211}
\end{figure*}

The channel maps of the lines listed in Table~\ref{table:lines+intensities} are presented in Fig.~\ref{fig:bit-hcop} to \ref{fig:bit-so2211}.    

\section{Comparison of different initial velocities}
\begin{figure*}
    \centering
    \includegraphics[width=0.85\textwidth]{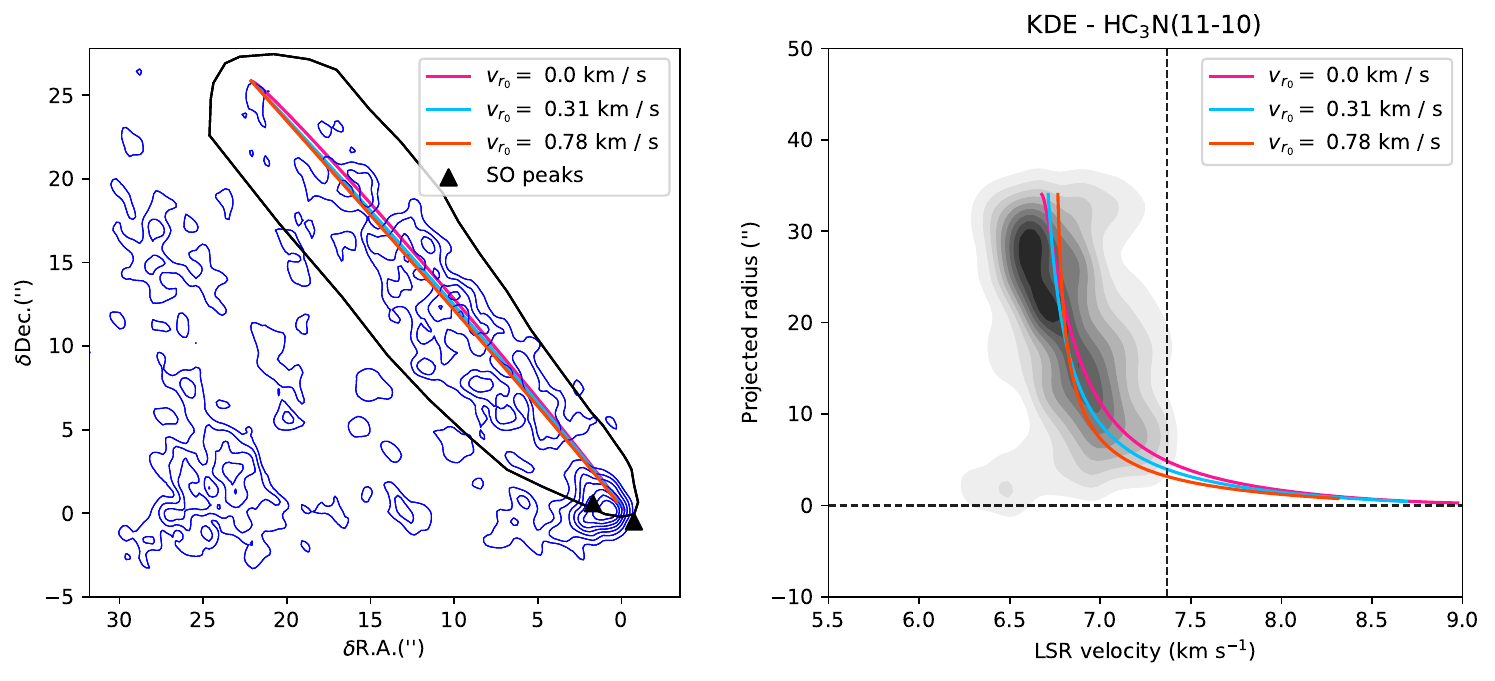}
    \caption{Streamline model of the \ce{HC3N} blue-shifted emission for the set of parameters $(r_0,~\theta_0,~\varphi_0)=$ (5000~au, 41\degr, $-170$\degr) around the $(\mathcal{R}_z)$ axis, with $v_{r,0}=0$~km~s$^{-1}$ (pink), $v_{r,0}=0.31$~km~s$^{-1}$ (blue) and $v_{r,0}=0.78$~km~s$^{-1}$ (orange). See Sect.~\ref{sec:streamline-model-HC3N} for more details. \textbf{Left:} Theoretical projected trajectories from the streamline model overlayed on the \ce{HC3N} blue-shifted emission and the two brightest peaks of the SO emission of Fig.~\ref{fig:streamer-shocks}. The contour levels are 7$\sigma$ to 19$\sigma$ by 2$\sigma$ steps where $\sigma=0.89$ mJy beam$^{-1}$ km~s$^{-1}$. The black contour delimits the mask used to compute the KDE. \textbf{Right:} Theoretical line of sight velocities profiles from the streamline model overlayed on the KDE of the \ce{HC3N} blue-shifted emission. The vertical dotted line corresponds to the disk $v_\text{LSR}=7.37$~km~s$^{-1}$, the horizontal to the 0\arcsec~offset.}
    \label{fig:streamline-compare-vel}
\end{figure*}

Figure~\ref{fig:streamline-compare-vel} compares the streamline model for \ce{HC3N} with the three different initial velocities used (see Sect.~\ref{sec:streamline-model-HC3N}).

\section{Herschel dust map}
\begin{figure*}
	\centering
	\includegraphics[width=0.7\linewidth]{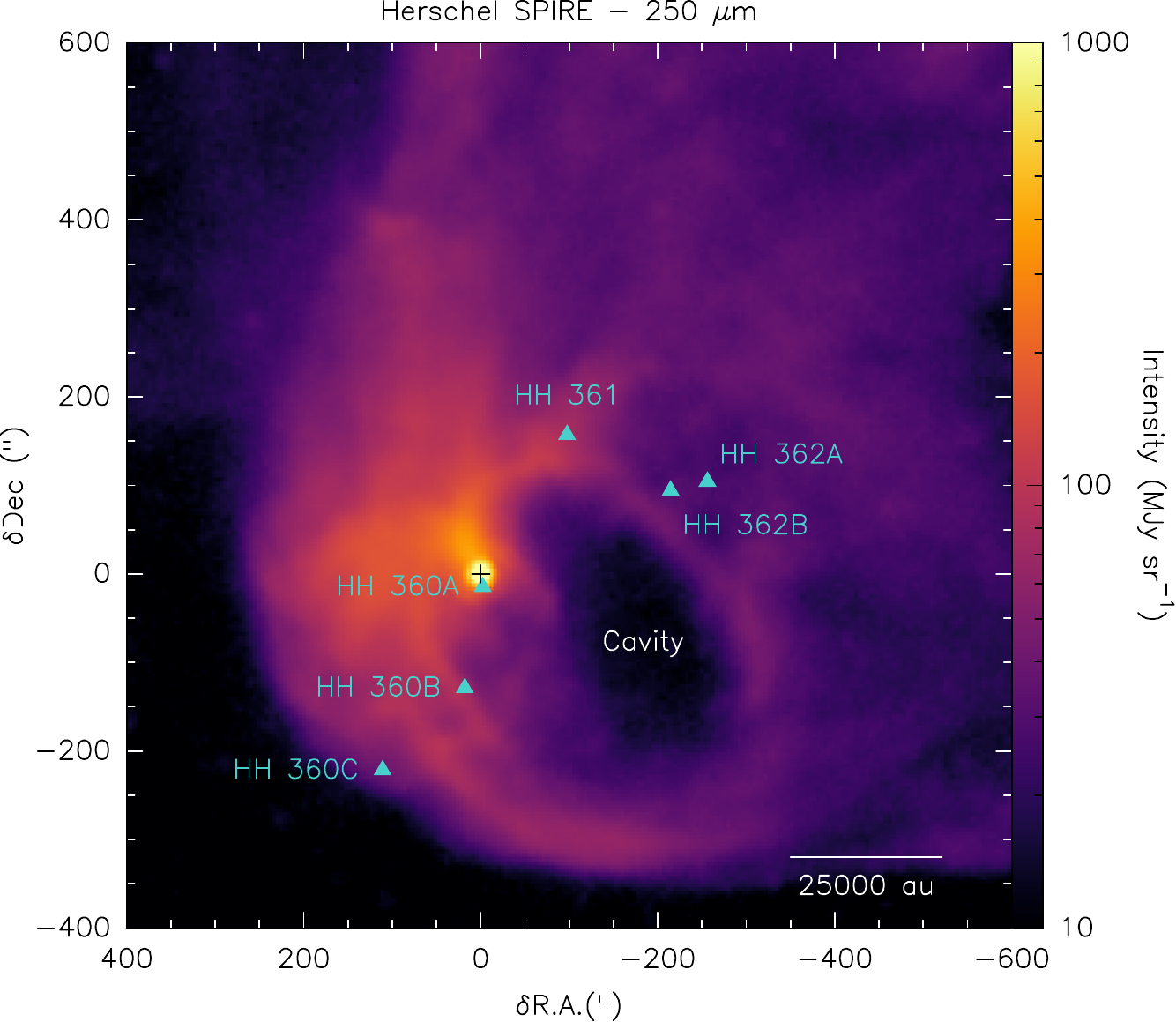}
	\caption{Herschel SPIRE dust map at $250~\mu m$ from the Herschel Key Program Guaranteed Time (KPGT, PI: P. Andre). The positions of the HH objects, represented as blue triangles, come from \citet{gomez1997}. The color scale is stretched by the log function and saturated on the $10-1000$~MJy~sr$^{-1}$ range. The cross denotes \object{L1489 IRS} position in (0\arcsec, 0\arcsec). The scale bar in the bottom right corner indicates 25000~au.\\}
	\label{fig:herschel-HH}
\end{figure*}

Figure~\ref{fig:herschel-HH} shows the Herschel/SPIRE map at $250~\mu m$ from the Herschel Key Program Guaranteed Time (KPGT, PI: P. Andre), where we overlaid the HH objects identified in \citet{gomez1997}.

\section{\texttt{TIPSY} fitting fraction and deviation}
\begin{figure*}
    \centering
    \includegraphics[width=\linewidth]{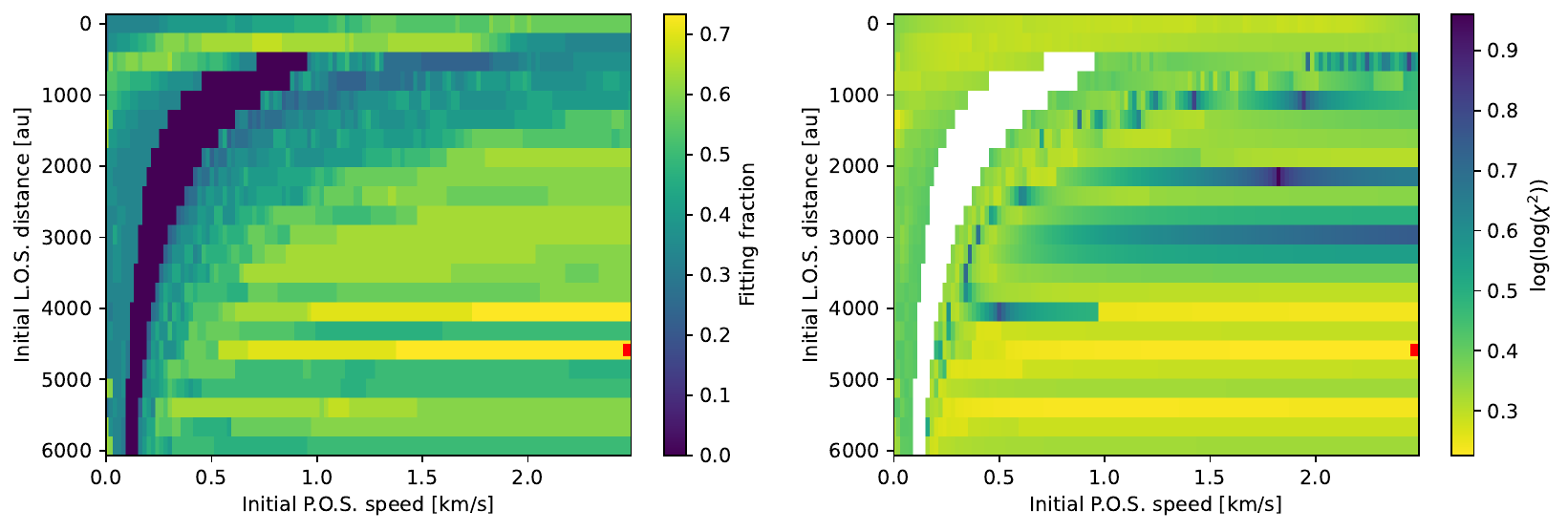}
    \caption{Quality of the \texttt{TIPSY} fit for the CS masked PPV cube (see Sect.~\ref{sec:tipsy-cs}). The red square indicates the best-fit model. \textbf{Left:} Fitting fraction of function of the initial distance on the line of sight (L.O.S) and the initial speed in the plane of sky (P.O.S.). \textbf{Right:} Like the left panel but for the deviation with $\log(\log(\;\chi^2))$ to enhance the contrast.}
    \label{fig:tipsy-cs-fit-chi2}
\end{figure*}

\begin{figure*}
    \centering
    \includegraphics[width=\linewidth]{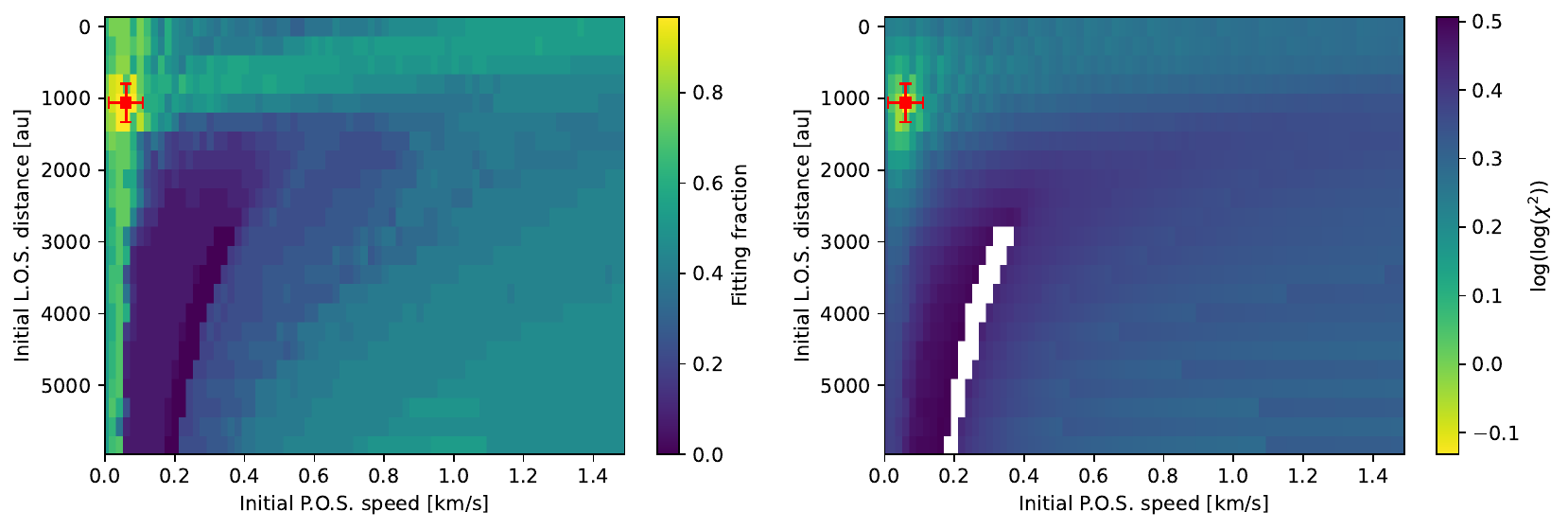}
    \caption{Same as Fig.~\ref{fig:tipsy-cs-fit-chi2} but for the \ce{HC3N} masked PPV cube (see Sect.~\ref{sec:tipsy-hc3n}).}
    \label{fig:tipsy-hc3n-fit-chi2}
\end{figure*}

Figures~\ref{fig:tipsy-cs-fit-chi2} (for CS, see Sect.~\ref{sec:tipsy-cs}) and \ref{fig:tipsy-hc3n-fit-chi2} (for \ce{HC3N}, see Sect.~\ref{sec:tipsy-hc3n}) present the 2D space of parameters explored by \texttt{TIPSY} models, that is initial distance along the line of sight and initial speed in the plane of sky, and their associated fitting fraction and $\chi^2$ deviation. The higher the fitting fraction is, the better the observations are reproduced, as it is defined as the fraction of observed values consistent with the theoretical infall trajectory (i.e., within observations uncertainties). The lower the $\chi^2$ deviation is, the better the fit is, as it is computed from observations nominal values. The combination of the two identifies the best-fit model (displayed with the red square). It is all the more reliable if it falls within a cluster of good parameters, like for \ce{HC3N} (see Fig.~\ref{fig:tipsy-hc3n-fit-chi2}) and not for CS (see Fig.~\ref{fig:tipsy-cs-fit-chi2}). Uncertainties of the best-fit model are calculated from good enough models, that means having a fitting fraction above 0.9. The absence of uncertainties (like for CS, see Fig.~\ref{fig:tipsy-cs-fit-chi2}) reflects the low reliability of the model.

\end{appendix}
\end{document}